\documentclass{aa} 
\usepackage{txfonts, newtxtext,newtxmath, bm, amsmath, amssymb, booktabs, array, enumerate, multirow, booktabs, ulem, threeparttable, soul, color, graphicx, extarrows} 
\usepackage[T1]{fontenc}
\usepackage[colorlinks=true, linkcolor=blue, citecolor=blue, urlcolor=blue]{hyperref} 
\usepackage{orcidlink}
\usepackage[switch]{lineno}

\DeclareRobustCommand{\VAN}[3]{#2}
\let\VANthebibliography\thebibliography
\def\thebibliography{\DeclareRobustCommand{\VAN}[3]{##3}\VANthebibliography}

\newcommand{\DOA}{Department of Astronomy, School of Physics, Peking University, Beijing 100871, China}
\newcommand{\KIAA}{Kavli Institute for Astronomy and Astrophysics, Peking University, Beijing 100871, China}
\newcommand{\MU}{School of Physics and Astronomy, Monash University, Clayton Victoria 3800, Australia}
\newcommand{\OzGrav}{OzGrav: The ARC Centre of Excellence for Gravitational Wave Discovery, Australia}
\newcommand{\HK}{Department of Physics, The University of Hong Kong, Pokfulam Road, Hong Kong, People's Republic of China}
\newcommand{\UNLV}{The Nevada Center for Astrophysics and Department of Physics and Astronomy, University of Nevada, Las Vegas, NV 89154, USA}
\newcommand{\HKIAA}{The Hong Kong Institute for Astronomy and Astrophysics, The University of Hong Kong, Pokfulam Road, Hong Kong, People's Republic of China}
\newcommand{\NAOC}{National Astronomical Observatories, Chinese Academy of Sciences, Beijing 100012, China}

\begin{document} 

\title{Mini-supernovae from white dwarf--neutron star mergers: Viewing-angle-dependent spectra and lightcurves}

\titlerunning{Viewing-angle dependence of WD--NS mini-supernovae}
\authorrunning{Kang et al.}


\author{Yacheng Kang\inst{1, 2,\thanks{email: yckang@stu.pku.edu.cn}}
		\orcidlink{0000-0001-7402-4927}
		\and
		Jin-Ping Zhu\inst{3, 4, 5, 6, \thanks{email: jpzhu@hku.hk}}
		\orcidlink{0000-0002-9195-4904}
		\and
		Lijing Shao\inst{2, 7,\thanks{email: lshao@pku.edu.cn}}
		\orcidlink{0000-0002-1334-8853}
		\and
		Jiahang Zhong\inst{1, 2}
		\orcidlink{0009-0008-2673-1764}
		\and
		Jinghao Zhang\inst{1, 2}
		\orcidlink{0009-0002-1101-2798}
		\and\\
		Bing Zhang\inst{3, 4, 8}
		\orcidlink{0000-0002-9725-2524}
		}
		

\institute{\DOA \and \KIAA \and \HKIAA \and \HK \and  \MU \and \OzGrav \and \NAOC \and \UNLV}
           
\date{Received Sep. 1, 2025 / Accepted Sep. 1, 2025}


\abstract
{Unstable mass transfer may occur during white dwarf--neutron star (WD--NS) mergers, in which the WD can be tidally disrupted and form an accretion disk around the NS. Such an accretion disk can produce unbound wind ejecta, with synthesized $^{56}$Ni mixed in. Numerical simulations reveal that this unbound ejecta should be strongly polar-dominated, which may cause the following radioactive-powered thermal transient to be viewing-angle-dependent---an issue that has so far received limited investigation.}
{We investigate how the intrinsically non‐spherical geometry of WD--NS wind ejecta affects the viewing‐angle dependence of the thermal transients.}
{Using a two‐dimensional axisymmetric ejecta configuration and incorporating heating from the radioactive decay of $^{56}\mathrm{Ni}$, we employ a semi‐analytical discretization scheme to simulate the observed viewing‐angle‐dependent photospheric evolution, as well as the resulting spectra and lightcurves.}
{The observed photosphere evolves over time and depends strongly on the viewing angle: off‐axis observers can see deeper, hotter inner layers of the ejecta and larger projected photospheric areas compared to on-axis observers. For a fiducial WD--NS merger producing $0.3\,{\rm{M}}_\odot$ of ejecta and $0.01\,{\rm{M}}_\odot$ of synthesized $^{56}$Ni, the resulting peak optical absolute magnitudes of the transient span from $\simeq -12\,\mathrm{mag}$ along the polar direction to $\simeq -16\,\mathrm{mag}$ along the equatorial direction, corresponding to luminosities of $\sim 10^{40}$--$10^{42}\,\mathrm{erg\,s^{-1}}$. The typical peak timescales are expected to be $3$--$10\,\mathrm{d}$.}
{We for the first time explore the viewing-angle effect on WD--NS merger transients. Since their ejecta composition and energy sources resemble those of supernovae, yet WD--NS merger transients are dimmer and evolve more rapidly, we propose using ``mini-supernovae'' to describe the thermal emission following WD--NS mergers. Our study highlights the critical role of geometry in interpreting WD--NS mini-supernovae and motivates further exploration of their diversity in observation.}



\maketitle


\section{Introduction}
\label{ sec:intro }

Among the various types of double-compact-object (DCO) mergers, systems that contain at least one neutron star (NS) are of particular interest, especially as promising progenitors of diverse electromagnetic (EM) transients and as detectable gravitational-wave (GW) sources. This interest was greatly enhanced by the detection of the first GW binary neutron star (BNS) merger, GW170817 \citep{LIGOScientific:2017vwq}, accompanied by the short-duration gamma-ray burst (SGRB) GRB\,170817A \citep{LIGOScientific:2017zic, Goldstein:2017mmi, Savchenko:2017ffs, 2018NatCo...9..447Z} and the kilonova transient AT\,2017gfo \citep{Andreoni:2017ppd, Arcavi:2017xiz, Chornock:2017sdf, Coulter:2017wya, Cowperthwaite:2017dyu, TOROS:2017pqe, Drout:2017ijr, Evans:2017mmy, Hu:2017tlb, Kasliwal:2017ngb, Kilpatrick:2017mhz, Lipunov:2017dwd, McCully:2017lgx, Nicholl:2017ahq, Pian:2017gtc, Pozanenko:2017jrn, Shappee:2017zly, Smartt:2017fuw, DES:2017kbs, Tanvir:2017pws, Troja:2017nqp, J-GEM:2017tyx, Valenti:2017ngx}. Such a multimessenger detection of a BNS GW event and its associated EM counterparts has provided smoking-gun evidence that BNS mergers can drive relativistic jets to power SGRBs \citep{Paczynski:1986px, Eichler:1989ve, 1991AcA....41..257P, Narayan:1992iy, Zhang:2018ads} and eject neutron-rich outflows undergoing $r$-process nucleosynthesis to produce kilonova emissions \citep{Li:1998bw, Metzger:2010sy, Metzger:2019zeh}. Moreover, they offered a unique opportunity to probe dense-matter physics and test gravity in extreme conditions \citep{LIGOScientific:2017pwl, LIGOScientific:2018cki, LIGOScientific:2018dkp, Shao:2017gwu, 2019BAAS...51c.251S, Zhao:2019suc, Shao:2022koz}. Thus, GW170817 is widely regarded as a milestone event in astronomy and fundamental physics \citep{LIGOScientific:2017ync, Nakar:2019fza, Radice:2020ddv, Margutti:2020xbo, Bian:2021ini, Nicholl:2024ttg}. 

In addition to BNS mergers, neutron star--black hole (NS--BH) mergers are also expected to produce observable SGRBs and kilonovae \citep{Paczynski:1986px, Eichler:1989ve, 1991AcA....41..257P, Narayan:1992iy, Zhang:2018ads}. Unfortunately, although several NS--BH GW events and candidates were reported during the LIGO-Virgo-KAGRA (LVK) Collaboration's third and fourth observing runs \citep[O3 and O4;][]{abbott2021gwtc,LIGOScientific:2021qlt,KAGRA:2021vkt,LIGOScientific:2024elc}, follow-up campaigns have to date failed to identify any convincing EM counterparts \citep[e.g.,][]{Gomez:2019tuj, Andreoni:2019qgh, Ackley:2020qkz, Kawaguchi:2020osi, Han:2020qmn, Kyutoku:2020xka}, in particular, the expected kilonovae. Such non-detections could result from the large distances of these events, incomplete coverage of the large GW localization regions, or the fact that many NS--BH mergers are direct plunges, leaving little to no ejecta \citep[e.g.,][]{LIGOScientific:2021qlt, Zhu:2021ysz, Zhu2021, Zhu:2024cvt, Fragione2021, LIGOScientific:2024elc}.

In recent years, several kilonova-like transients have been unexpectedly discovered in association with long-duration GRBs (LGRBs), e.g., LGRB\,211211A \citep{Rastinejad:2022zbg, Yang:2022qmy, Troja:2022yya} and LGRB\,230307A \citep{JWST:2023jqa, 2024Natur.626..742Y, Gillanders:2024hcb}. These discoveries challenge the classical GRB classification, since phenomenological LGRBs are typically thought to originate from the core collapse of massive stars, which are usually accompanied by broad-lined type Ic supernovae \citep[SNe;][]{Woosley:1993wj, Paczynski:1997yg, Woosley:1998hk, MacFadyen:1998vz, Metzger:2010sy, Zhang:2018ads, Metzger:2019zeh}, rather than by kilonovae that are associated with DCO mergers. If these merger-driven LGRBs indeed arise from BNS or NS--BH mergers, the observed kilonova emission can be naturally interpreted; however, their long prompt-emission durations remain difficult to reconcile with the current understanding of jet production in such DCO mergers, where the accretion episodes that launch jets are expected to be relatively short \citep{Narayan:1992iy, Yuan:2012jy, Zhang:2018ads, Zhang:2024syc}. Taken together, these theoretical and observational tensions suggest that the merger-driven LGRBs may constitute a distinct population beyond the classical GRB classification \citep{Zhu:2024xtm, Kang:2025eoh, Wang:2024ems, Zhu:2025woo, Tan:2025jvl, Maccary:2025zus}. White dwarf--neutron star (WD--NS) mergers are increasingly investigated as a plausible progenitor scenario for merger-driven LGRBs, as they can potentially power long-duration relativistic jets, owing to the lower densities and extended free-fall timescales of WDs, while simultaneously producing optical follow-up transients \citep{Yang:2022qmy, Zhong:2023zwh, Zhong:2024alw, Wang:2024ijo, Peng:2021knv, Du:2024kst, Chen:2024arq,Zhang:2024syc, Liu:2025voz, Xu:2025ugj, Chrimes:2025mfy, Arunachalam:2025rnd, Chen:2025zpv}.

Despite growing interest, multimessenger investigations of WD--NS mergers have not yet reached the same level of depth as those of other types of DCO mergers in the literature. In fact, WD--NS binaries are likely the most common compact binaries aside from WD--WD binary systems \citep{Nelemans:2001hp, Toonen:2018njy}. Millions of WD--NS binaries are expected to exist in our Galaxy, and hundreds of known pulsars are already found in DCO systems with WD companions \citep[ATNF pulsar catalogue\footnote{\url{http://www.atnf.csiro.au/research/pulsar/psrcat/}};][]{Manchester:2004bp, Bobrick:2017rlo, 2025RAA....25a4003W}. Present population-synthesis studies estimate that the WD--NS merger rate lies in the range of $[8, 500]\,\mathrm{Myr}^{-1}$ per Milky-Way-like (MW-like) galaxy \citep{Nelemans:2001hp, Toonen:2018njy, Kaltenborn:2022vxf}\footnote{For comparison, the observed type Ia SN rate is approximately $4000\,\mathrm{Myr}^{-1}$ per MW-like galaxy \citep{Toonen:2018njy}. Based on GW observations reported in GWTC-3, the local BNS merger rate is estimated to be in the range of $[1, 30]\,\mathrm{Myr}^{-1}$ per MW-like galaxy \citep{KAGRA:2021duu}.}. A large fraction ($\sim60$--$80$\%) of WD--NS mergers are predicted to involve CO WDs, with most of the rest involving ONe WDs \citep{Toonen:2018njy, Kaltenborn:2022vxf}. Compared to BNSs and NS--BH systems, the intrinsically low density and extended structure of WDs impart unique characteristics to WD--NS mergers \citep{King:2006nw, Chattopadhyay:2007rx, Paschalidis:2011ez, Margalit:2016joe, Margalit:2016hlx}. These properties can trigger substantial early mass transfer and lead to the formation of a massive, extended accretion disk, ultimately producing polar-dominated unbound ejecta outflows, consisting primarily of original WD material and synthesized iron-group elements \citep{Metzger:2011zk, Fernandez:2012kh, Fernandez:2019eqo, Zenati:2018gcp, Zenati:2019glf, Bobrick:2021hho, Moran-Fraile:2023oui}. Although the wide dynamical ranges in time and length scales pose computational challenges for exploring the nature of the central remnant in WD--NS mergers\footnote{The existence of relativistic jets and associated high-energy emission in WD--NS mergers still remains uncertain and debated in the literature.}, many studies in the literature have nevertheless developed a broadly consistent merger scenario by focusing on the evolution of the accretion disk and the generation of unbound ejecta\footnote{Most simulations excise the central engine region by imposing an inner boundary.}. This picture can be summarized as follows:
\begin{enumerate}[(a)]
   \item Once a WD--NS binary is formed, it loses orbital energy through GW radiation, causing the orbit to shrink over time. When the binary separation decreases to the Roche-lobe overflow \citep[RLOF;][]{Eggleton:1983rx}, mass begins transferring from the WD to the NS companion.
   \item During the mass-transfer stage, the orbital separation may increase due to angular momentum conservation, whereas the WD expansion as it loses mass increases the minimal orbital separation required for RLOF, with these two effects competing. Many studies have shown that the stability of mass transfer depends sensitively on the WD-to-NS mass ratio $q = M_{\mathrm{WD}}/M_{\mathrm{NS}}$ \citep[e.g.,][]{1988ApJ...332..193V, Paschalidis:2009zz, Margalit:2016joe}. For $q \lesssim 0.5$, most low-mass He WD--NS binaries are expected to undergo stable mass transfer \citep{2022ApJ...930..134C}; otherwise, the process becomes unstable, leading to the rapid, runaway disruption of the WD on a dynamical timescale\footnote{\citet{Bobrick:2017rlo} argued that winds from the accretion stream can further destabilize the system, yielding a much lower critical value $q_{\text {crit }} \simeq 0.20$, which is consistent with results by \citet{Kaltenborn:2022vxf}.}. CO WD--NS and ONe WD--NS binaries are generally thought to belong to this unstable regime \citep{Bobrick:2017rlo, Kaltenborn:2022vxf}.
   \item For WD--NS binaries that may undergo an unstable mass-transfer phase, many (nuclear) hydrodynamical simulations have explored the evolution of the accretion disk and resulting wind outflows \citep{Fernandez:2012kh, Fernandez:2019eqo, Zenati:2018gcp, Zenati:2019glf, Bobrick:2021hho, Moran-Fraile:2023oui}. Most of the disk is expected to be radiatively inefficient and to evolve primarily through viscous processes \citep{Metzger:2011zk, Margalit:2016joe}. As material accretes to smaller radii, it reaches higher temperatures and undergoes burning that synthesizes progressively heavier elements. Combined viscous and nuclear heating drives substantial amounts of unbound wind ejecta, which are funneled almost vertically from the innermost regions of the disk, preferentially along the polar directions.
   \item The ejecta are composed mostly of the original WD material,  with some synthesized iron-group material, mainly $^{56}\mathrm{Ni}$, mixed in. As WDs are proton-rich, the outflows are not as neutron-rich as those from BNS or NS--BH mergers and therefore fail to synthesize substantial $r$-process elements \citep{Fernandez:2019eqo, Bobrick:2021hho, Kaltenborn:2022vxf, Liu:2025voz}. Finally, the radioactive decay of synthesized $^{56}\mathrm{Ni}$ can potentially power week-scale thermal transients following WD--NS mergers.
\end{enumerate}

Given this picture, the unbound ejecta from WD--NS mergers undergoing unstable mass transfer are expected to be intrinsically anisotropic, which would produce highly viewing-angle-dependent EM counterparts. In this work, we therefore explore how this asymmetric geometry affects the viewing‐angle dependence of the thermal emission from WD--NS mergers along different lines of sight, a question that has so far received limited attention. Using a two-dimensional (2D) asymmetric ejecta configuration and incorporating radioactive heating from the decay of $^{56}\mathrm{Ni}$, we employ a semi‐analytical discretization method to model the viewing-angle-dependent photosphere and its temperature evolution. In our proposed model with fiducial ejecta mass ($M_{\rm ej} \simeq 0.3\,\mathrm{M}_\odot$) and $^{56}\mathrm{Ni}$ mass ($M_{\rm Ni} \simeq 0.01\,\mathrm{M}_\odot$), we find that the peak optical timescales of WD--NS merger transients are $\sim3$--$10\,{\rm d}$. Their peak optical absolute magnitudes are strongly viewing-angle-dependent, ranging from ${\simeq -12\,\mathrm{mag}}$ in the polar direction to ${\simeq -16\,\mathrm{mag}}$ in the equatorial direction, corresponding to peak optical luminosities of $\simeq 10^{40}$--$10^{42}\,\mathrm{erg} \, \mathrm{s}^{-1}$. Given that their ejecta composition and energy sources are similar to those of SNe, while WD--NS merger transients are dimmer and evolve on shorter timescales, we adopt the term ``mini-supernovae'' (mSNe) to describe the thermal emission following WD--NS mergers. Our findings also highlight that the asymmetry of the wind ejecta has a significant impact on the photosphere evolution, temperature evolution, emergent spectra, and multiband lightcurves observed from different viewing angles. This work thus motivates further investigations into the diversity and observable signatures of WD--NS merger counterparts.

The structure of this paper is as follows. Sect.~\ref{ sec:ejecta_profile } describes the semi-analytical construction of the WD--NS ejecta distribution and dynamics. Sect.~\ref{ sec:temperature } and Sect.~\ref{ sec:photosphere } present the evolution of the ejecta temperature and the photospheric structure for different viewing angles, respectively. The viewing-angle-dependent emergent spectra and lightcurves of WD--NS mSNe are shown in Sect.~\ref{ sec:lc_spec }. A detailed discussion is provided in Sect.~\ref{ sec:discussion }. Finally, our conclusions are summarized in Sect.~\ref{ sec:conclu }.


\section{Ejecta distribution and dynamics}
\label{ sec:ejecta_profile }

In this section, we focus on constructing semi-analytical models for WD--NS binaries that may undergo an unstable mass-transfer phase. The total mass of the disk wind ejecta ($M_\mathrm{ej}$) and the amount of $^{56}\mathrm{Ni}$ it contains ($M_\mathrm{Ni}$) are parameterized as fractions of the initial WD mass ($M_\mathrm{WD}$), expressed as ${M_\mathrm{ej} = f_{\mathrm{ej}} M_{\mathrm{WD}}}$ and $M_\mathrm{Ni} = f_{\mathrm{Ni}} M_{\mathrm{WD}}$, respectively. We adopt $f_{\mathrm{ej}} = 0.3$ and ${f_{\mathrm{Ni}} = 0.01}$ for a classical merger of a $1.4\,\mathrm{M}_{\odot}$ NS and a $1\,\mathrm{M}_{\odot}$ WD (i.e., ${M_{\mathrm{WD}} = 1\,\mathrm{M}_{\odot}}$) as our fiducial model\footnote{Systems with different internal compositions can exhibit different ejecta properties, e.g., ONe WD--NS mergers are expected to eject more mass and yield higher $^{56}$Ni than CO WD--NS mergers \citep{Fernandez:2019eqo, Bobrick:2021hho}.}. The choices of these fiducial parameters are supported by previous studies \citep[see e.g.,][]{Fernandez:2019eqo, Zenati:2019glf, Kaltenborn:2022vxf, Moran-Fraile:2023oui}. The impact of varying $f_{\mathrm{ej}}$ and $f_{\mathrm{Ni}}$ will be discussed further in Sect.~\ref{ sec:discussion }. 

In this work, the distribution and dynamics of disk outflows in WD--NS mergers are primarily informed by the post-merger simulations of \citet{Fernandez:2019eqo}\footnote{Note that the main characteristics of wind dynamics and distribution in \citet{Fernandez:2019eqo} are consistent with those reported in several other studies, including \citet{Fernandez:2012kh, Zenati:2018gcp, Moran-Fraile:2023oui}.}. Specifically, we adopt a 2D axisymmetric model of the wind ejecta in spherical coordinates, assuming no dependence on the azimuthal angle $\varphi$. Given a latitudinal mass distribution function $F(\theta)$ in spherical coordinates ($r$, $\theta$), the total mass of the wind ejecta within a latitudinal angle $\theta$ ($0 \leqslant \theta \leqslant \frac{\pi}{2}$) can be expressed as
\begin{equation}
\begin{aligned} m_{\mathrm{ej}}
& = \int_{0}^{\theta} \int_{0}^{2\pi} F(\theta^\prime) \sin \theta^\prime \it{d} \varphi d \theta^\prime \\
& = 2\pi \int_{0}^{\theta} F(\theta^\prime) \sin \theta^\prime d \theta^\prime \\ 
& = -2\pi \int_{0}^{\theta} F(\theta^\prime) d(\cos \theta^\prime)\,.\end{aligned}
\label{ eq:mass_w_theta }
\end{equation}
Owing to the symmetry of the mass distribution about the equatorial plane, we extend the distribution to the other hemisphere by adopting ${m_{\mathrm{ej}}(\theta) = m_{\mathrm{ej}}(\pi - \theta)}$ for ${\frac{\pi}{2} < \theta \leqslant \pi}$. Introducing a new variable $\mu \equiv \cos \theta \in [0, 1]$, Eq.~(\ref{ eq:mass_w_theta }) can then be rewritten as 
\begin{equation}
m_{\mathrm{ej}} = -2\pi \int_{1}^{\mu} G(\mu^\prime ) d \mu^\prime \,.
\label{ eq:mass_w_mu }
\end{equation}
To derive a more realistic expression for $G(\mu)$ or $F(\theta)$, we adopt a latitudinal distribution of the wind ejecta closely matching the angular profile shown in Fig.~2 of \citet{Fernandez:2019eqo}, which can be well approximated by
\begin{equation}
\mathrm{log}_{10} \left[G(\mu)\right] \simeq 2\mu + C_0 \,,
\label{ eq:angle_distri }
\end{equation}
or equivalently,
\begin{equation}
G(\mu) = A\times10^{2\mu} \,,
\label{ eq:angle_distri_2 }
\end{equation}
where $C_0$ and $A$ are constants to be determined. The normalization condition requires
\begin{equation}
\frac{M_{\mathrm{ej}}}{2} = -2\pi \int_{1}^{0} G(\mu^\prime ) d \mu^\prime = 2 \pi A \int_{0}^{1} 10^{2\mu^\prime} d \mu^\prime =\frac{99 \pi}{\ln 10} A\,.
\label{ eq:norm_condition }
\end{equation}
Therefore, we obtain $A = \frac{\ln 10}{198 \pi} M_{\mathrm{ej}}$, which allows Eq.~(\ref{ eq:angle_distri_2 }) to be rewritten as: 
\begin{equation}
G(\mu) = \frac{\ln 10 }{198\pi}M_{\mathrm{ej}} \times 10^{2\mu}\,.
\label{ eq:G_w }
\end{equation}
Substituting $G(\mu)$ into Eq.~(\ref{ eq:mass_w_theta }) and Eq.~(\ref{ eq:mass_w_mu }), the mass within a polar angle $\theta$ can be expressed as
\begin{equation}
\begin{aligned}	
m_{\mathrm{ej}} 
& = -\frac{\ln 10 }{99}M_{\mathrm{ej}} \int_{1}^{\mu} 10^{2\mu^\prime} d \mu^\prime \\
& = \frac{\ln 10 }{99}M_{\mathrm{ej}} \int_{0}^{\theta} 10^{2\cos\theta^\prime}\sin\theta^\prime d \theta^\prime\,. 
\end{aligned}
\label{ eq:mass_w_mu_2 }
\end{equation}
Therefore, the angle-dependent mass per unit solid angle in a given direction ($\theta, \varphi$) can be written as
\begin{equation}
\frac{dm_\mathrm{ej}}{d\Omega} \equiv \frac{dm_\mathrm{ej}}{\sin \theta d\theta d \varphi} = C_1 \times 10^{2 \cos \theta} \,,
\label{ eq:mass_per_solid_angle }
\end{equation}
where the coefficient is defined as ${C_1 \equiv \frac{\ln 10 }{198 \pi}M_{\mathrm{ej}} \simeq 0.0037M_{\mathrm{ej}}}$.

Previous numerical simulations suggested that the faster-moving ejecta is primarily distributed along the polar direction \citep[see e.g., ][]{Fernandez:2012kh, Zenati:2018gcp, Fernandez:2019eqo, Moran-Fraile:2023oui}, which motivates the assumption of an ellipsoidal velocity profile (hereafter referred to as the ``v-profile''), described by
\begin{equation}
\left(\frac{\varv_{\rm max} \cos \theta}{\varv_{z, {\rm max}}}\right)^2 + \left(\frac{\varv_{\rm max} \sin \theta}{\varv_{e, {\rm max}}}\right)^2 = 1 \,,
\label{ eq:v_profile }
\end{equation}
where $\varv_{\rm max}$ denotes the maximum velocity of the wind ejecta in the direction specified by $\theta$. We adopt $\varv_{z, {\rm max}} = 0.1 c$ and $\varv_{e, {\rm max}} = 0.01c$ as the axial and equatorial velocities at $\theta = 0$ and $\theta = \pi/2$, respectively, where $c$ is the speed of light. These parameter values here are supported by previous studies \citep[see e.g., ][]{Metzger:2011zk, Margalit:2016joe, Fernandez:2019eqo, Zenati:2018gcp, Bobrick:2021hho, Moran-Fraile:2023oui}. Based on Eq.~(\ref{ eq:v_profile }), the angular dependence of the maximum velocity is then given by
\begin{equation}
\varv_{\rm max}(\theta) = \frac{{\varv}_{z, {\rm max}}{\varv}_{e, {\rm max}}} {\sqrt{\varv_{z, {\rm max}}^2\sin^2\theta + \varv_{e, {\rm max}}^2\cos^2\theta}} \,.
\label{ eq:v^max_w }
\end{equation}

Assuming a homologous expansion \citep[i.e., $r = \varv t$;][]{Zenati:2019glf, Kaltenborn:2022vxf} and a flat distribution of wind ejecta mass per velocity bin in a given ($\theta, \varphi$) direction \citep{Kyutoku:2015gda, Kawaguchi:2016ana}, we obtain an alternative normalization condition,
\begin{equation}
\begin{aligned}	
\frac{d \mathit{m}_\mathrm{ej}}{d\Omega} = \int_{0}^{r_{\rm max}} \rho r^2 dr = \int_{0}^{\varv_{\rm max}} \rho \varv^2 t^3 d\varv \,,	
\end{aligned}	
\label{ eq:v_normalization }
\end{equation}
where $\rho(\varv,\theta,t)$ is the mass density within the wind ejecta. Combining it with Eq.~(\ref{ eq:mass_per_solid_angle }), we find
\begin{equation}
\frac{d m_{\mathrm{ej}}}{ d\Omega d \varv} = \rho \varv^2 t^3 \simeq C_1 \times 10^{2 \cos \theta}/\varv_{\rm max} \,.
\label{ eq:dm_dv }
\end{equation}
Note that $\varv_{\rm max}$ is derived in Eq.~(\ref{ eq:v^max_w }). Using the above equations, we can then express the density $\rho(\varv,\theta,t)$ as
\begin{equation}
\rho(\varv,\theta,t) = C_1 \times 10^{2\cos\theta} /\varv_{\mathrm{max}} \varv^{2} {t}^{3}\,.
\label{ eq:rho }
\end{equation}

\begin{figure*}[htbp]
  \centering
  \includegraphics[width=5.4cm]{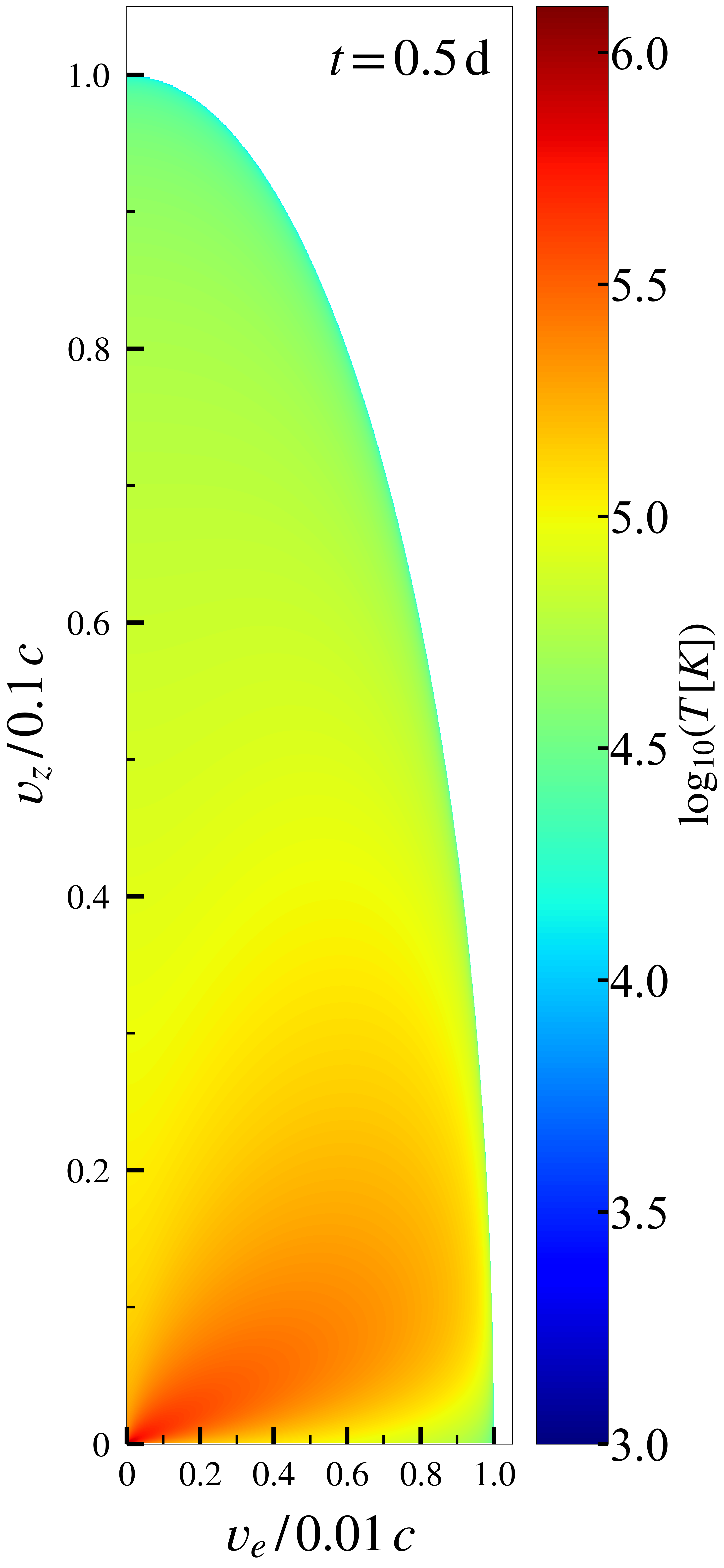} 
  \hspace{0.4cm}
  \includegraphics[width=5.4cm]{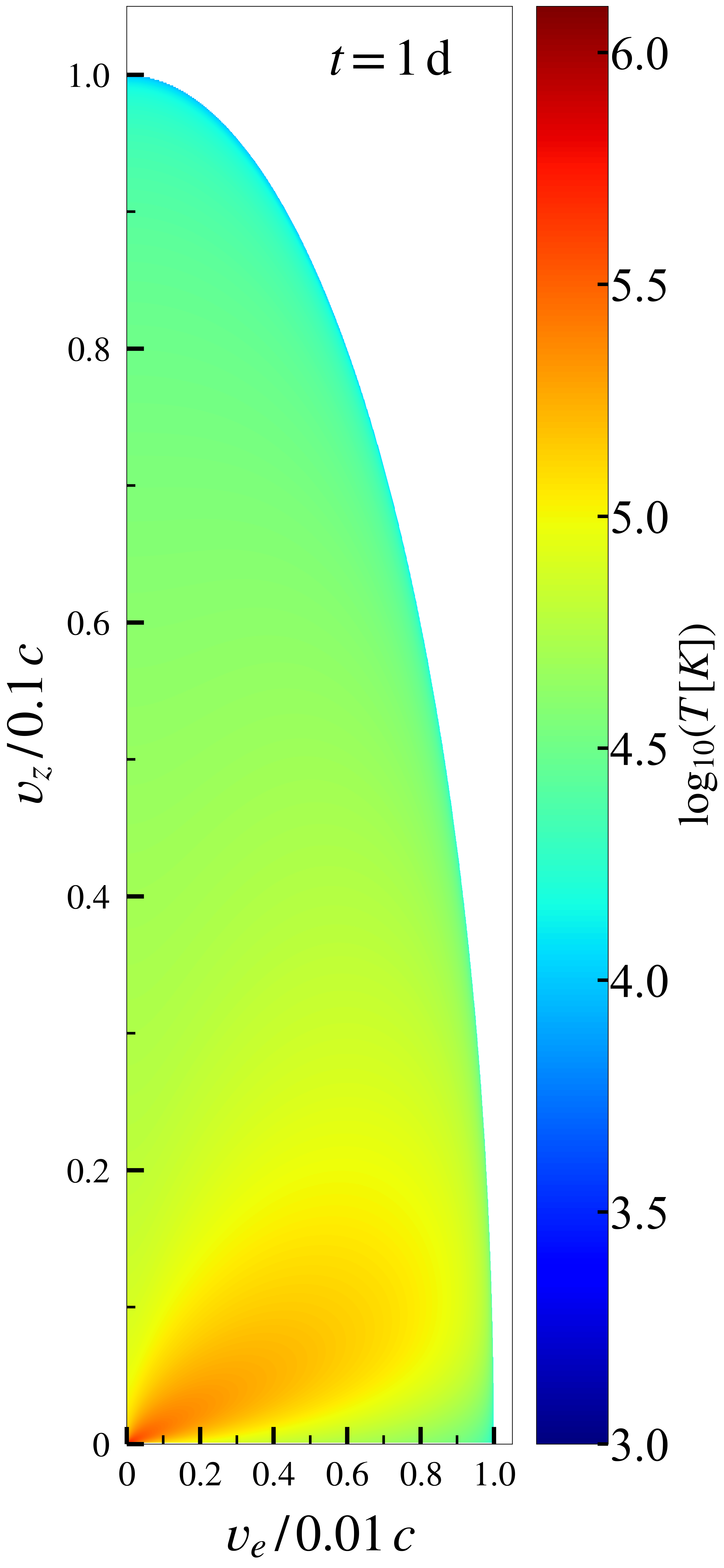} 
  \hspace{0.4cm}
  \includegraphics[width=5.4cm]{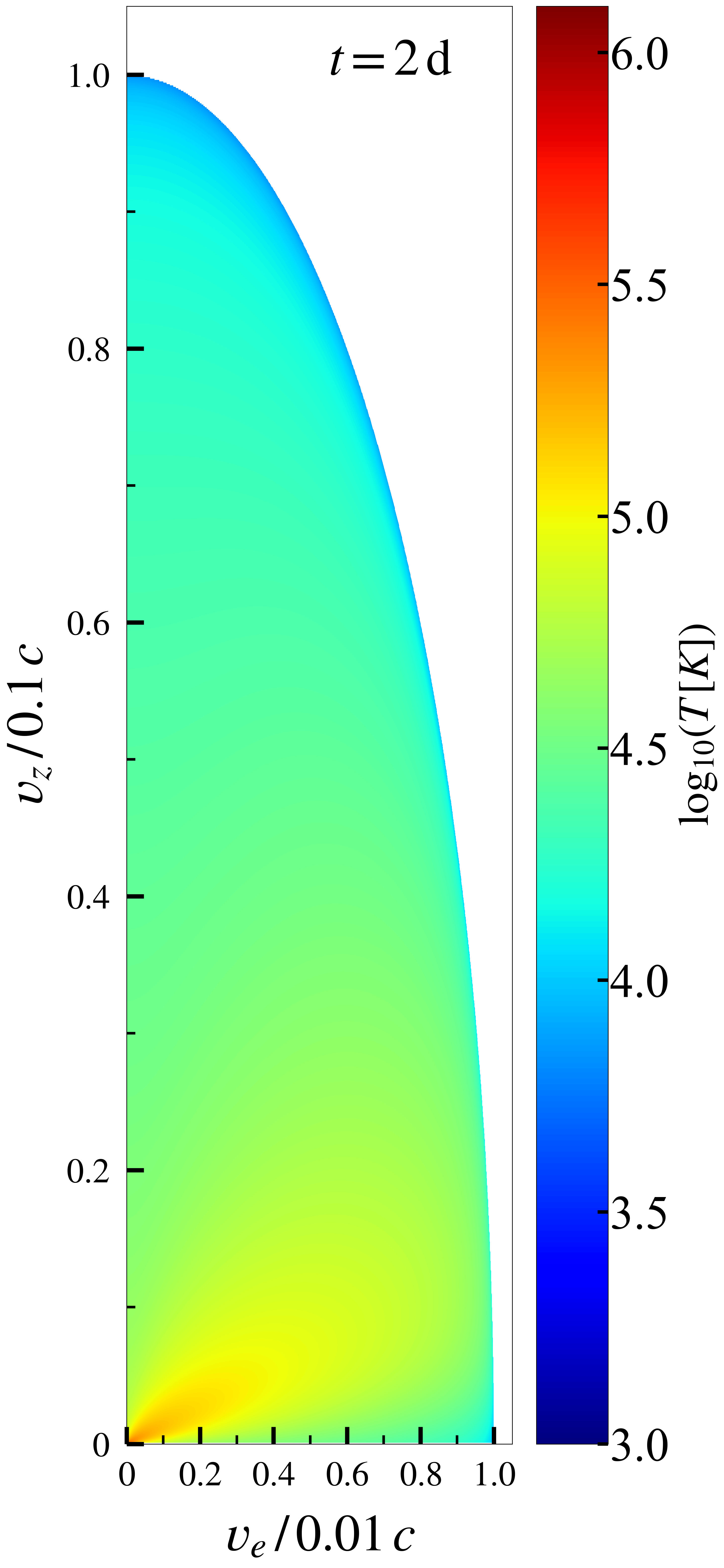} 
  \\[0.3cm]
  \includegraphics[width=5.4cm]{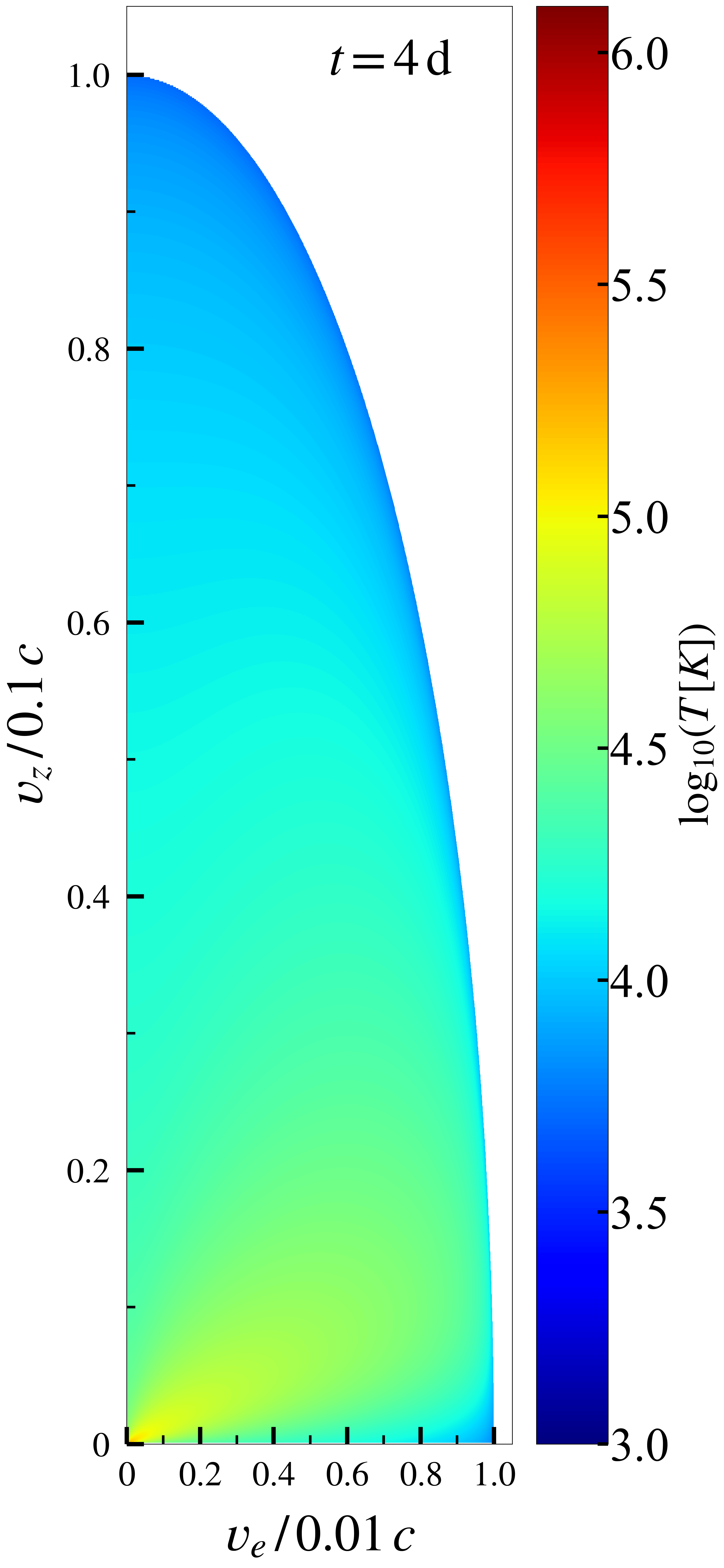} 
  \hspace{0.4cm}
  \includegraphics[width=5.4cm]{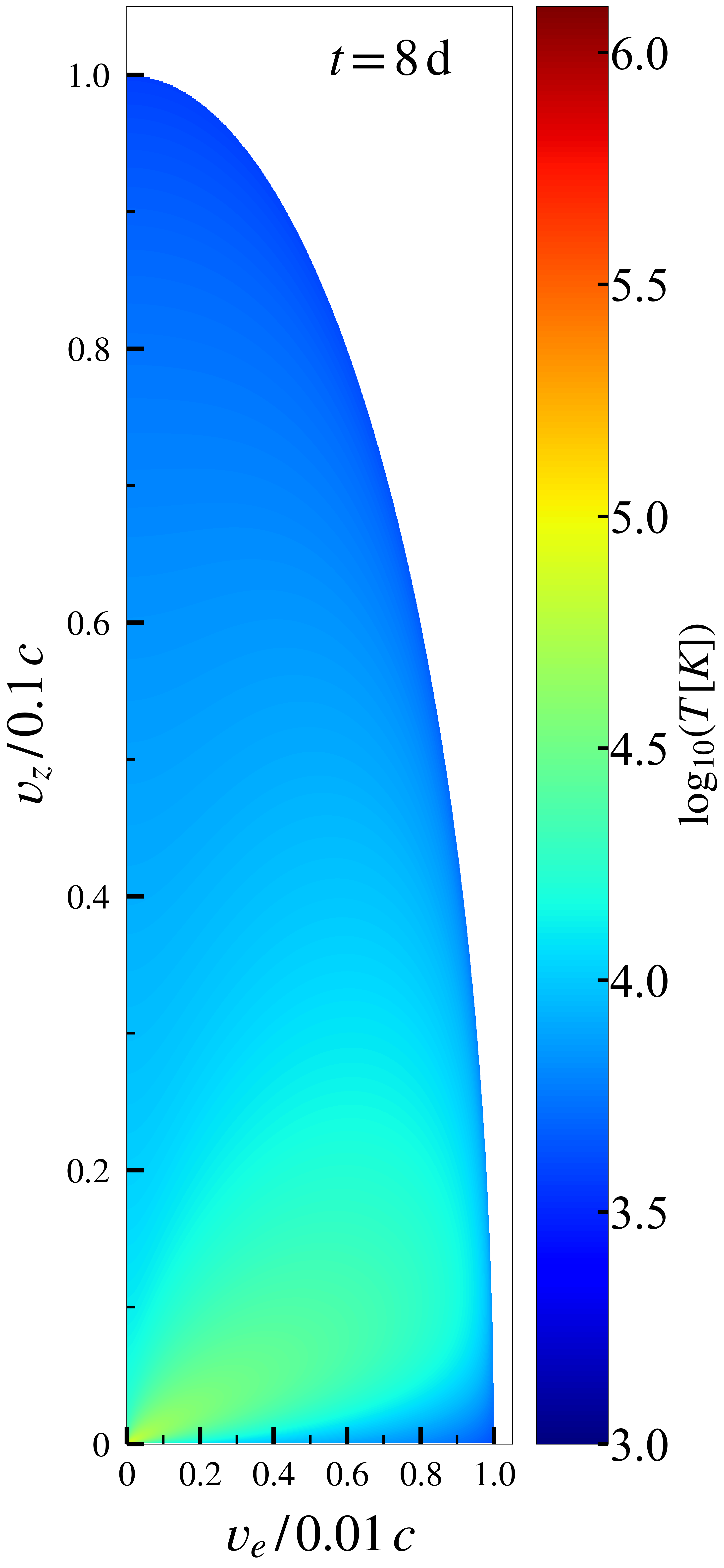} 
  \hspace{0.4cm}
  \includegraphics[width=5.4cm]{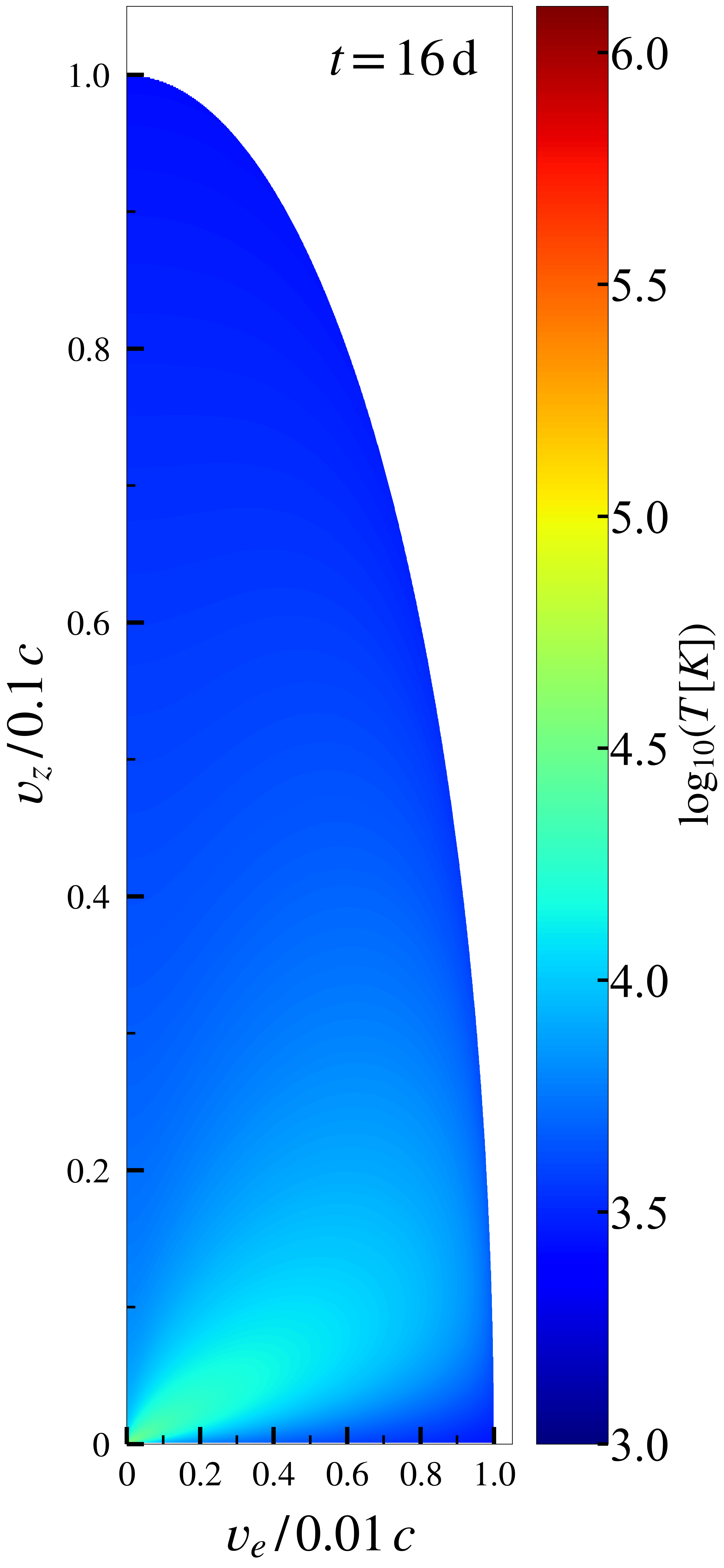}
  \caption{Sectional drawings of the temperature evolution within the ejecta at $t = 0.5\,\mathrm{d}$, $1\,\mathrm{d}$, $2\,\mathrm{d}$, $4\,\mathrm{d}$, $8\,\mathrm{d}$, and $16\,\mathrm{d}$ following the WD--NS merger of our fiducial model. Here, $\varv_{z}$ and $\varv_e$ denote the ejecta velocities along the polar axis ($\theta = 0$) and the equatorial plane ($\theta = \pi/2$), respectively. The colour scale indicates the logarithmic temperature. Given the 2D axisymmetry of the system in spherical coordinates, the temperature structure can be extended to the entire ejecta.}
  \label{ fig:temperature }
\end{figure*}


\section{Evolution of the ejecta temperature}
\label{ sec:temperature }

From the ejecta v-profile and mass density distribution given by Eqs. (\ref{ eq:v^max_w }) and (\ref{ eq:rho }), respectively, we derive the optical depth along the radial direction at a given time $\it{t}$:
\begin{equation}
\begin{aligned}	
\tau(\varv, \it{t}) 
& \simeq \int_{\varv}^{\varv_{\rm max}} \kappa_{\mathrm{ej}} \rho \it{t} d \varv' \\
& = C_1 \kappa_{\mathrm{ej}} t^{-2} \int_{\varv}^{\varv_{\rm max}}  \frac{10^{2\cos\theta}}{\varv_{\rm max}\varv'^{2}} \mathit{d\varv}' \\
& = C_1 \kappa_{\mathrm{ej}} \frac{10^{2\cos\theta} (\varv_{\rm max} - \varv)}{\varv_{\mathrm{\rm max}}^{2} \varv \it t^{2}} \,,
\end{aligned}
\label{ eq:tau_rad }
\end{equation}
where $\kappa_{\mathrm{ej}}$ denotes the opacity of the ejecta, for which we adopt a constant\footnote{For simplicity, we adopt a gray-opacity approximation to describe the overall thermal evolution and lightcurve morphology of the WD--NS merger ejecta. Nevertheless, frequency-dependent line opacities, as explored in various radiative-transfer studies of SNe and kilonovae, can alter the detailed spectral features and colors, especially through UV line blanketing by species such as C and O at late times \citep{Pinto:1999ah, Kasen:2013xka, Tanaka:2019iqp, Magee:2021nau}. A detailed investigation of these effects in WD--NS thermal transients is beyond the scope of this work and should be addressed in future studies.} value of ${\kappa_{\mathrm{ej}}=0.2 \mathrm{~cm}^{2} \mathrm{~g}^{-1}}$, consistent with previous works \citep{Nagy:2018vxl, Liu:2025voz}. Following \citet{Zhu:2020inc}, we can further locate the photon diffusion surface by the condition $ {(\varv_{\rm max} - \varv_{\rm diff}) \it{t} \simeq ct/\tau(\varv_{\rm diff}, \mathit{t})}$, i.e., ${\tau(\varv_{\rm diff}, \mathit{t}) \simeq \it{c} / (\varv_{\rm max} - \varv_{\rm diff})}$. Below this surface, photons are trapped within the ejecta, since their diffusion velocity is slower than the expansion velocity of the ejecta. Similarly, we adopt ${\tau(\varv_{\rm phot}) \simeq 2/3}$ as the criterion to define the ejecta's photosphere. Both $\varv_{\rm diff}$ and $\varv_{\rm phot}$ can thus be determined from Eq.~(\ref{ eq:tau_rad }) at any given time.  

During the WD--NS merger, we assume that the ejecta are fully mixed, consistent with the assumptions adopted or tested in previous simulation studies \citep[see e.g.,][]{Fernandez:2012kh, Fernandez:2019eqo, Bobrick:2021hho}. Given the values of $\varv_{\rm diff}$ and $\varv_{\rm phot}$, the total emissivity per unit area in a given ($\theta, \varphi$) direction, denoted as $\mathcal{E}$, can be expressed as
\begin{equation}
\begin{aligned}	
\mathcal{E}(t)
& \simeq \int_{\varv_{\rm diff}}^{\varv_{\rm max}} \frac{f_{\mathrm{Ni}} \dot{\epsilon}(t)  \left(1 - \it e^{-\tau_{\gamma}}\right) d m_{\mathrm{ej}}}{f_{\mathrm{ej}} \varv_{\rm phot}^{2} \mathit{t^2} d\Omega}\\
& = \frac{f_{\mathrm{Ni}} \dot{\epsilon}(t) t \int_{\varv_{\rm diff}}^{\varv_{\rm max}} \rho \varv^{2}  \left(1 - \it e^{-\tau_{\gamma}}\right) \mathit{d} \varv}{f_{\mathrm{ej}} \rm \varv_{phot}^{2}}\\
& = C_1 \times 10^{2\cos\theta} \frac{f_{\mathrm{Ni}} \dot{\epsilon}(t) \int_{\varv_{\rm diff}}^{\varv_{\rm max}}  \left(1 - \it e^{-\tau_{\gamma}}\right) \mathit{d} \varv}{f_{\mathrm{ej}}  \varv_{\rm phot}^{2} \varv_{\rm max} \it t^2}  \,,
\end{aligned}
\label{ eq:emissivity }
\end{equation}
where $\dot{\epsilon}(t)$ is the radioactive heating rate per unit mass. Since WD--NS mergers are not expected to yield substantial $r$-process material \citep{Fernandez:2019eqo, Bobrick:2021hho, Kaltenborn:2022vxf, Moran-Fraile:2023oui, Chen:2024acv, Liu:2025voz}, we assume that the heating is entirely powered by the decay of $^{56}$Ni and $^{56}$Co \citep[i.e., the decay chain ${ }^{56} \mathrm{Ni} \rightarrow{ }^{56} \mathrm{Co} \rightarrow{ }^{56} \mathrm{Fe}$;][]{1994ApJS...92..527N, Arnett:1996ev}
\begin{equation}
\dot{\epsilon}(t) \simeq \left(3.3 e^{-t / t_{\mathrm{Ni}}} + 0.73 e^{-t / t_{\mathrm{Co}}}\right) \times 10^{10}\,\mathrm{erg}\,\mathrm{s}^{-1}\,\mathrm{g}^{-1} \,,
\label{ eq:e_injec }
\end{equation}
where $t_{\mathrm{Ni}} = 8.8\,\mathrm{d}$ and $t_{\mathrm{Co}} = 111.3\,\mathrm{d}$ are the mean lifetimes of $^{56}$Ni and $^{56}$Co, respectively. Our adopted heating rate corresponds to the portion of the radioactive decay energy that is deposited into the ejecta, and therefore already excludes the energy carried away by neutrinos. Note that we include a correction factor of $1 - \it e^{-\tau_{\gamma}}$ in Eq.~(\ref{ eq:emissivity }) to incorporate the effect of gamma-ray leakage \citep{1997ApJ...491..375C, Chatzopoulos:2009uz, Chatzopoulos:2011vj, Wang:2014cqa}. Similar to Eq.~(\ref{ eq:tau_rad }), $\tau_{\gamma}$ can be approximated to
\begin{equation}	
\tau_{\gamma}(\varv, \it{t}) \simeq C_{\rm 1} \kappa_{\gamma} \frac{\rm 10^{2\cos\theta} (\varv_{\rm max} - \varv)}{\varv_{\mathrm{\rm max}}^{2} \varv \it t^{2}} \,,
\label{ eq:tau_gamma }
\end{equation}
where we adopt $\kappa_{\gamma} = 0.4 \mathrm{~cm}^{2} \mathrm{~g}^{-1}$ for the gamma-ray opacity, a value considered appropriate for CO-dominated ejecta \citep[see e.g., ][]{Nicholl:2017vde, Zhu:2025zpk}. The corresponding $\gamma$-ray optical depth decreases with time as $\tau_{\gamma} \propto t^{-2}$. Adopting ${\varv \sim \varv_{\rm max} / 2}$, we estimate the characteristic transparency timescale along the polar or equatorial direction as,
\begin{equation}	
t_{\gamma, \mathrm{trans }} \simeq 36 \mathrm{~d}\left(\frac{\kappa_{\gamma}}{0.4 \mathrm{~cm}^{2} \mathrm{~g}^{-1}}\right)^{1 / 2}\left(\frac{M_{\mathrm{ej}}}{0.3\,\mathrm{M}_{\odot}}\right)^{1 / 2}\,,
\label{ eq:t_gamma }
\end{equation}
at which $\tau_{\gamma} \simeq 1$. Beyond this epoch, the $\gamma$-ray deposition efficiency $1 - \it e^{-\tau_{\gamma}}$ decreases rapidly, leading to significant $\gamma$-ray leakage. Once the ejecta become optically thin to $\gamma$-rays ($\tau_\gamma \lesssim 1$), the local energy deposition by positrons, which carry a few percent of the total decay energy, may start to dominate the late-time heating \citep{1994ApJS...92..527N, 1997ApJ...491..375C, Wang:2014cqa}. A more detailed radiative-transfer treatment including positron transport should be explored in future studies.

Once $\mathcal{E}$ is known, the effective temperature of the ejecta can be calculated via the Stefan-Boltzmann law,
\begin{equation}
T_{\rm eff}(t)=\left(\frac{\mathcal{E}}{\sigma_{\mathrm{SB}}}\right)^{1 / 4} \,,
\label{ eq:effec_t }
\end{equation}
where $\sigma_{\mathrm{SB}}$ is the Steffan-Boltzmann constant. Under the Eddington approximation \citep{1979rpa..book.....R}\footnote{The Eddington approximation assumes local thermodynamic equilibrium (LTE). In the expanding ejecta of WD--NS mergers, LTE gradually breaks down as the optical depth decreases to $\tau \lesssim 1$, typically after $\sim 30\,\mathrm{d}$ for our fiducial ejecta parameters. Beyond this epoch, non-LTE effects, such as deviations from blackbody spectra and non-thermal excitation, are expected to become important \citep[see e.g.,][]{Pinto:1999ah, Kromer:2009zq, Tanaka:2013ana}.}, the thermal temperature at each velocity shell can then be written as
\begin{equation}
T(\varv, \it t)=T_{\mathrm{\rm eff}}(t)\left\{\rm \frac{3}{4}\left[\tau(\varv, \it t) + \rm \frac{2}{3}\right]\right\}^{\rm 1 / 4} \,.
\label{ eq:temperature }
\end{equation}

We present the temporal evolution of the ejecta temperature profiles following WD--NS merger of our fiducial model in Fig.~\ref{ fig:temperature }. Owing to the 2D axisymmetry of the system in spherical coordinates, we show only central slices of the ejecta perpendicular to the equatorial plane. As shown in Fig.~\ref{ fig:temperature }, at early times ($\simeq 0.5\,\mathrm{d}$) after the merger, the ejecta reaches a temperature as high as $\sim 10^6\,\mathrm{K}$, but cools rapidly due to significant expansion. By $\simeq 2\,\mathrm{d}$, the temperature of most of the ejecta has decreased to $\sim 10^{4.5}\,\mathrm{K}$. At later times, as the ejecta expands homologously, the relative expansion rate (i.e., the fractional increase in radius per unit time) diminishes, leading to a reduced cooling efficiency. Consequently, it takes nearly $\simeq 16\,\mathrm{d}$ for the bulk of the ejecta to cool down to $\sim 10^{3.5}\,\mathrm{K}$.


\section{Photosphere evolution in the observer frame}
\label{ sec:photosphere }

In order to obtain the viewing-angle-dependent photosphere and photospheric temperature evolution as seen by a distant observer, we follow the discretization scheme outlined in \citet{Zhu:2020inc}, with minor adjustments to suit our WD--NS ejecta model\footnote{We recommend that readers interested in more detailed procedural descriptions can be referred to Appendix~B of \citet{Zhu:2020inc}.}. We first construct a mesh grid plane in velocity space, defined by sharing the same origin and $x$-axis as the local velocity coordinate system of the ejecta, and being perpendicular to the line of sight. Each mesh grid point is represented by a velocity vector in the local velocity coordinate system, denoted as $\vec{\varv}_{\text{mesh}}^{i j} = \left(\varv_{\text{mesh},x}^{i j}, \varv_{\text{mesh},y}^{i j}, \varv_{\text{mesh},z}^{i j}\right)$, where the superscripts ($i$ and $j$) represent the IDs of the points at the mesh grid. For each point on the mesh grid plane, we further construct a grid line within the ejecta along the line-of-sight direction, denoted as $p_{k}^{ij}$ with the subscript $k$ representing the ID of the points at each grid line. Note that $p_{1}^{ij}$ is defined as the first intersection point of the ejecta profile with the line of sight, i.e., the point at which the observer sees the smallest optical depth. Given the mass density distribution~(see Eq.~(\ref{ eq:rho })), the optical depth between $p^{ij}_k$ and $p^{ij}_1$ can be expressed as
\begin{equation}
\begin{aligned}	
\tau^{ij}(p_k, t) 
& = \int^{p_1^{ij}}_{p_k^{ij}} \kappa_{\mathrm{ej}} \rho t d p \\
& = C_1 \kappa_{\mathrm{ej}} t^{-2} \int^{p_1^{ij}}_{p_k^{ij}}  \frac{10^{2\cos\theta}}{\varv^{2}\varv_{\rm max}} \mathit{dp}\,, \hspace{1cm} k > 1\,,
\end{aligned}
\label{ eq:tau }
\end{equation}
where
\begin{equation}
\begin{aligned}	
&\varv^{2}
=  p^{2} + \varv_{\text{mesh}}^{i j}{}^{2} + 2 p \left(\varv_{\text{mesh},y}^{i j} \sin \theta_{\text{view}} + \varv_{\text{mesh},z}^{i j}\cos \theta_{\text{view}}\right) \,,
\\
&\varv_{\text{mesh}}^{i j}{}^{2} 
 = \varv_{\text{mesh},x}^{i j}{}^{2} + \varv_{\text{mesh},y}^{i j}{}^{2} + \varv_{\text{mesh},z}^{i j}{}^{2}  \,,
\end{aligned}
\label{ eq:tau_integrand }
\end{equation}
and $p$ denotes the distance in velocity space from the mesh grid plane to the intersection point\footnote{In the local coordinate system of the ejecta, we define a vector $\vec{p}$ pointing from $\vec{\varv}_{\text{mesh}}^{i j}$ to the intersection point. Its direction determines the sign of $p$: $p < 0$ if $\vec{p}$ points along the line of sight, and $p > 0$ otherwise.}. The latitudinal viewing angle is denoted by $\theta_{\text{view}}$, and the angle used in the density function is calculated via ${\cos\theta = |\varv_{z}/\varv|}$\footnote{As mentioned in Sect.~\ref{ sec:ejecta_profile }, we consider $\cos \theta \in [0, 1]$ in this work.}, where ${\varv_{z} = \varv_{\text{mesh},z}^{i j}+p \cos \theta_{\mathrm{view}}}$. In this work, we adopt ${\tau^{ij} \simeq 2/3}$ as the criterion for the photospheric surface along the line of sight. The observed location of the photosphere ($p_{\text{phot}}^{ij}$) at time $t$ is therefore given by
\begin{equation}
\tau^{ij}(p_{\text {phot }}^{i j}, t) 
= C_1 \kappa_{\mathrm{ej}} t^{-2} \int^{p_1}_{p_{\text {phot }}^{i j}}  \frac{10^{2\cos\theta}}{\varv^{2}\varv_{\rm max}} \mathit{dp} \simeq 2/3\,.
\label{ eq:tau_obs }
\end{equation}
By combining the photospheric location with the ejecta temperature structure shown in Fig.~\ref{ fig:temperature }, one can obtain the thermal temperature distribution across the entire photospheric surface in the observer frame, denoted as $T^{i j}_{\rm phot}$ throughout this work.

\begin{figure*}[htbp]
  \centering
  \includegraphics[width=8cm]{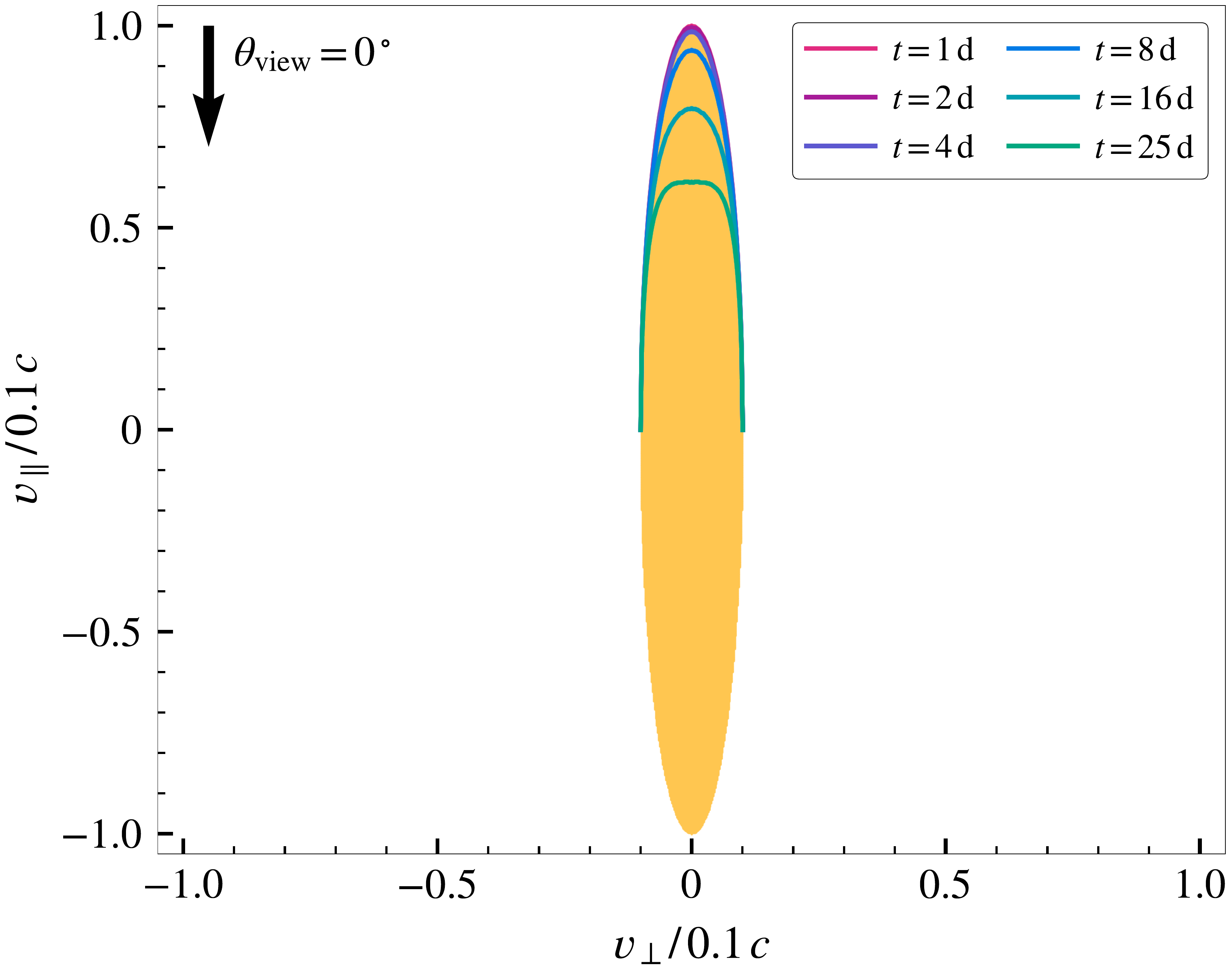}
  \hspace{1cm}
  \includegraphics[width=8cm]{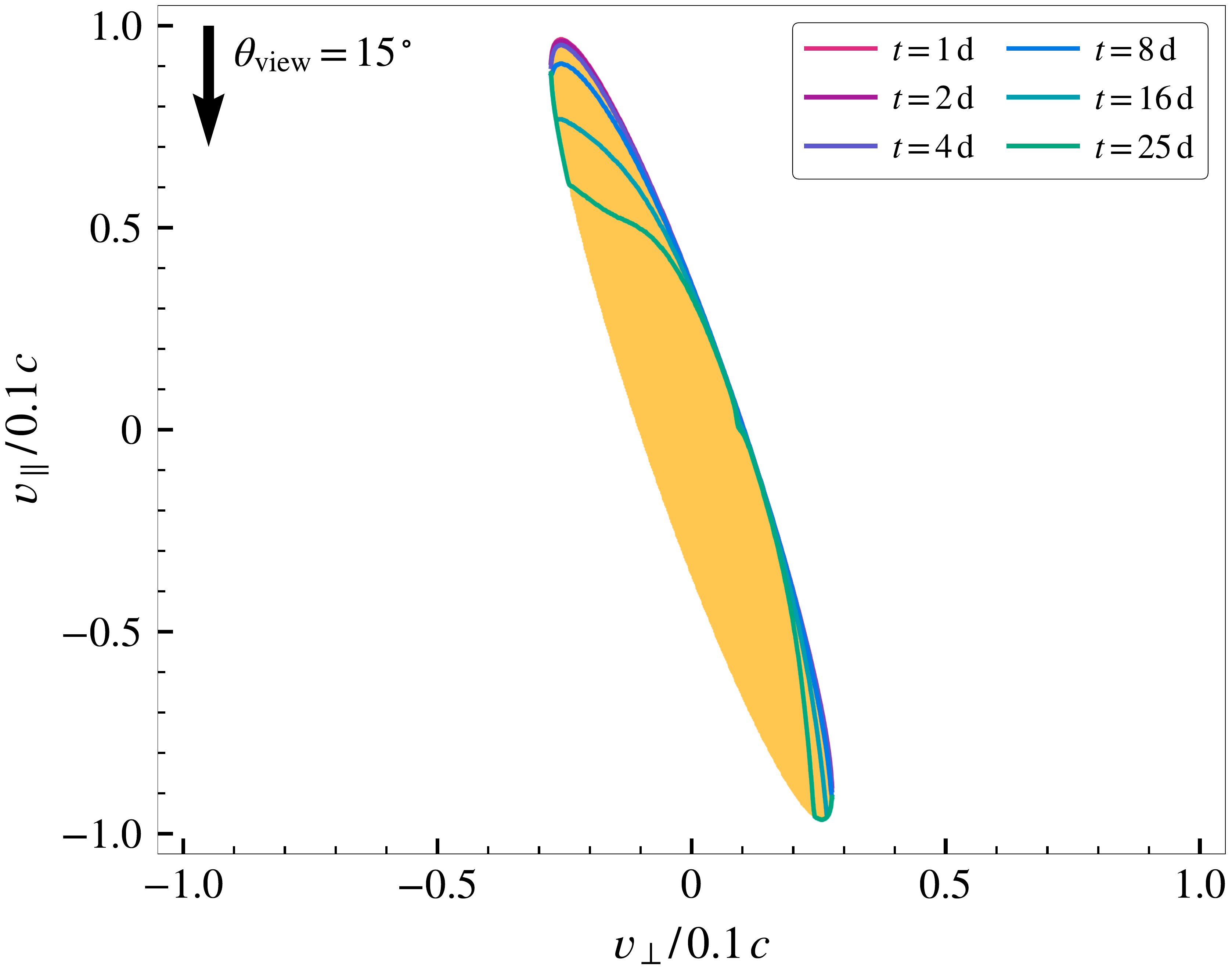}
  
  \vspace{0.3cm}
  
  \includegraphics[width=8cm]{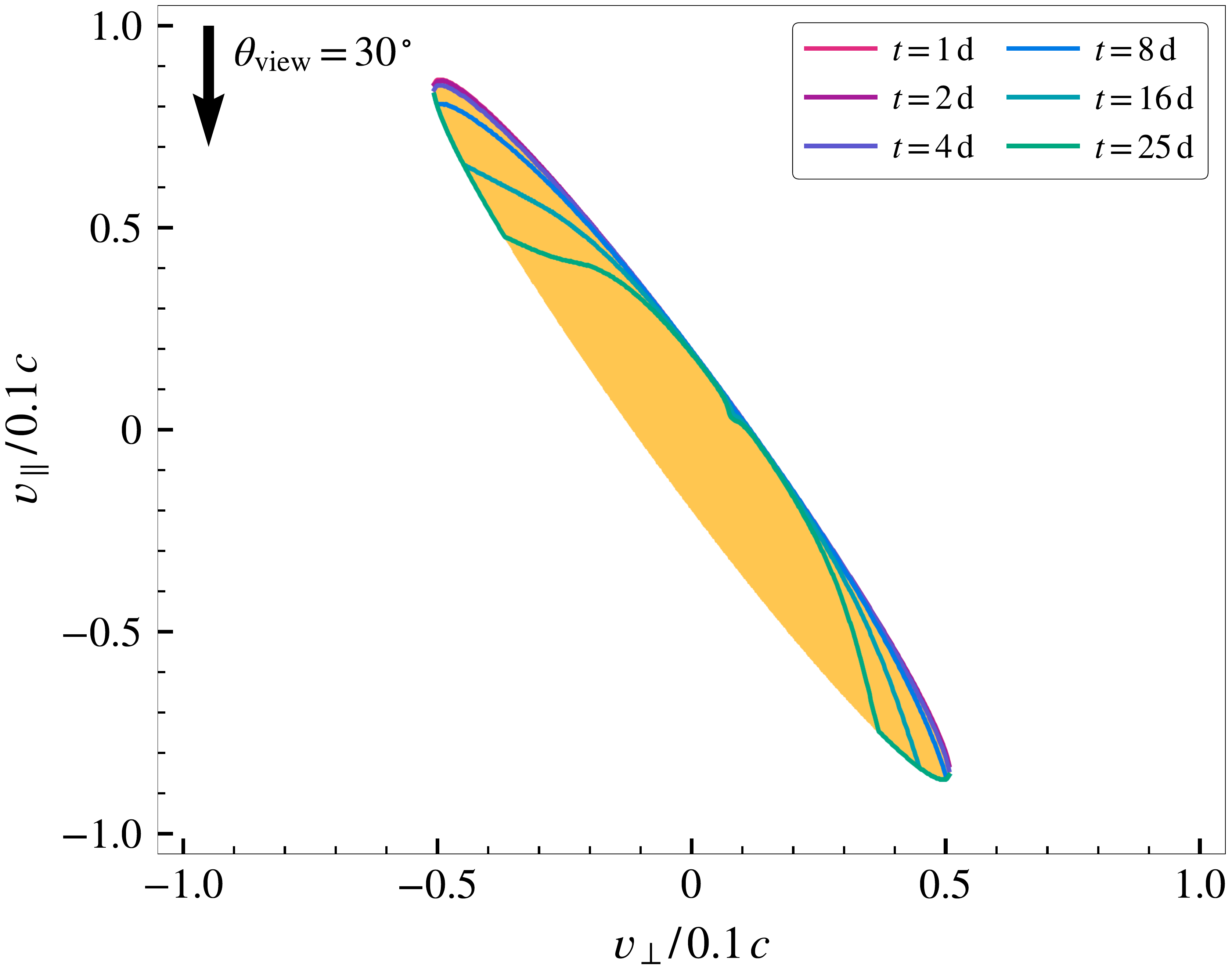}
  \hspace{1cm}
  \includegraphics[width=8cm]{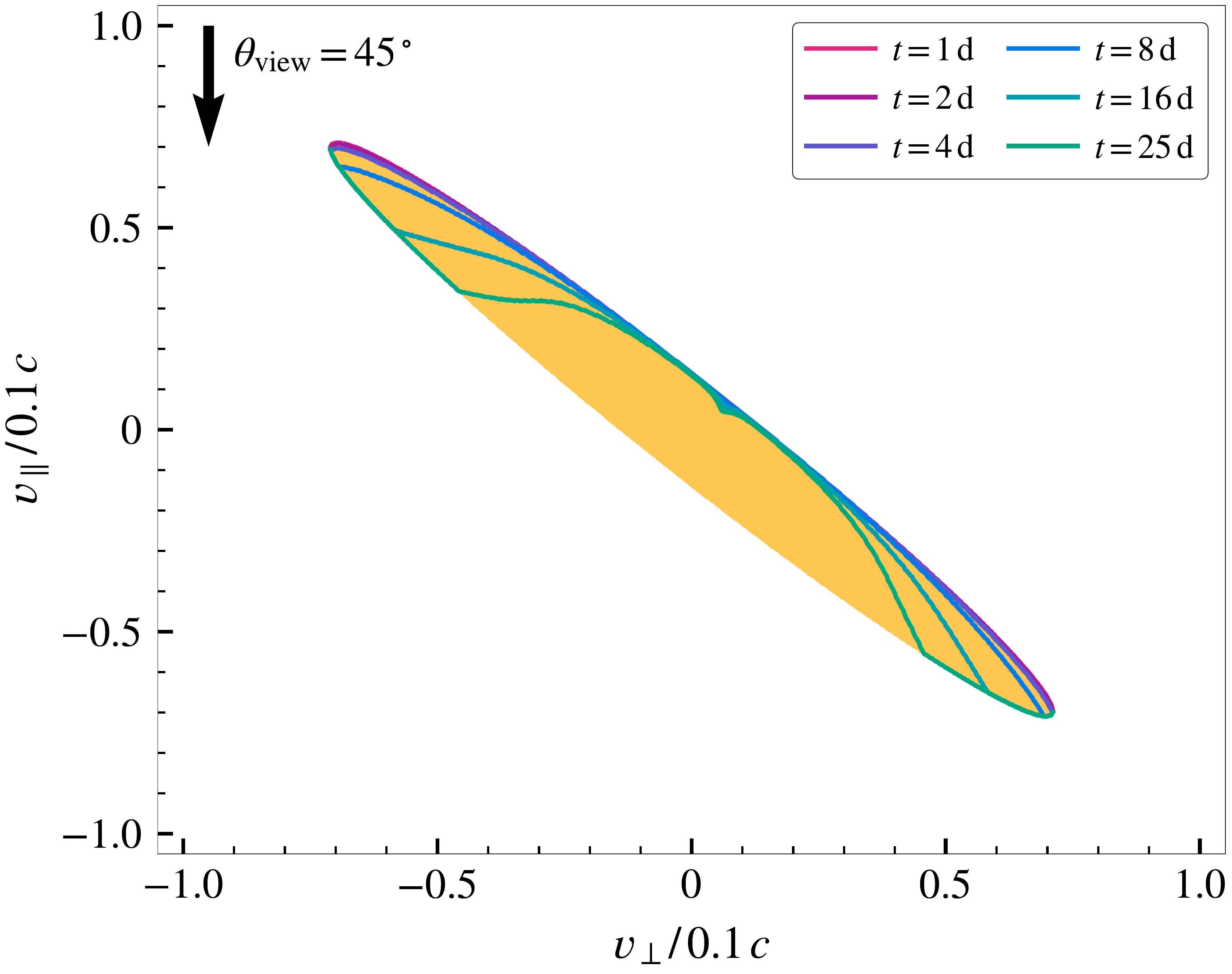}
  
  \vspace{0.3cm}
  
  \includegraphics[width=8cm]{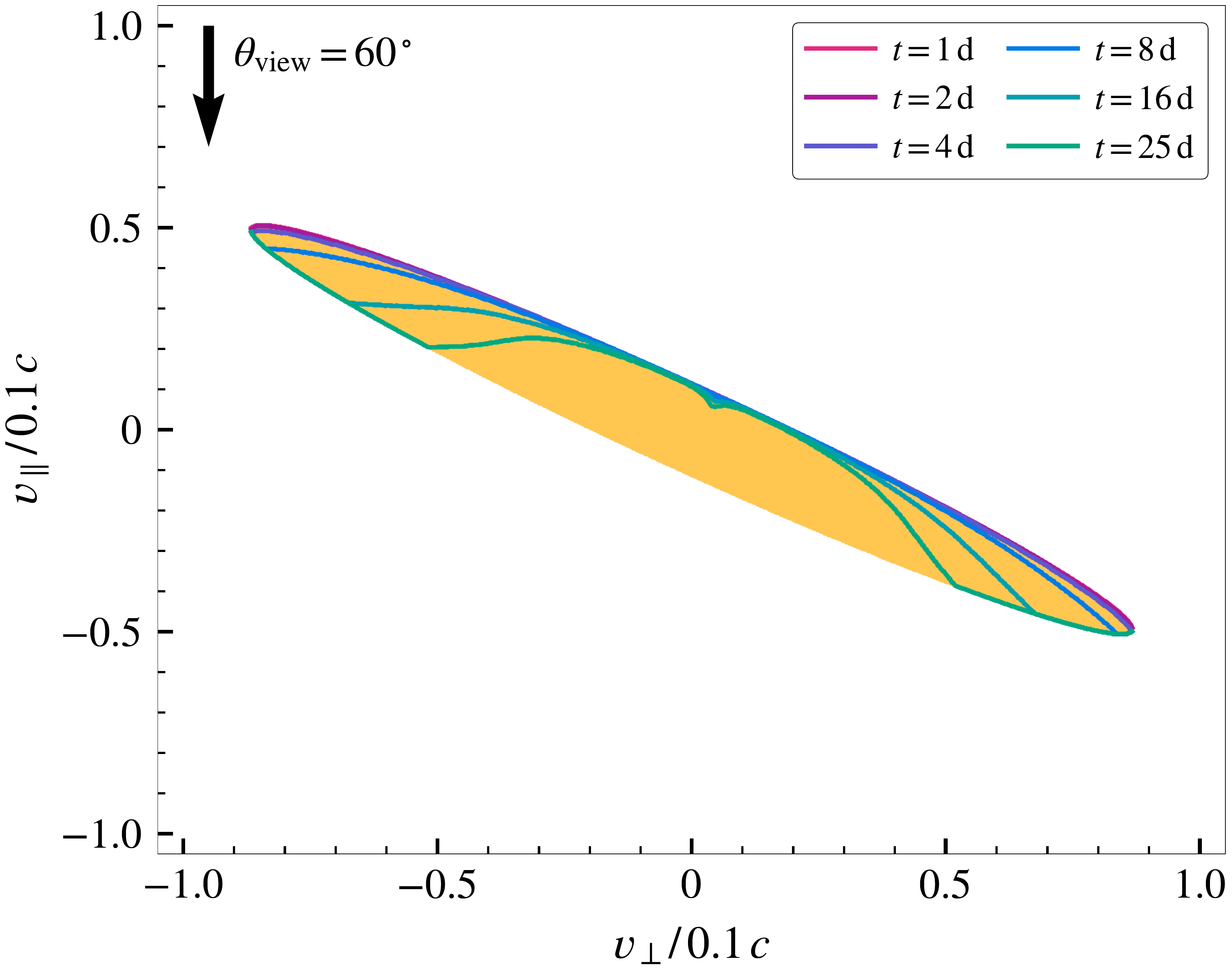}
  \hspace{1cm}
  \includegraphics[width=8cm]{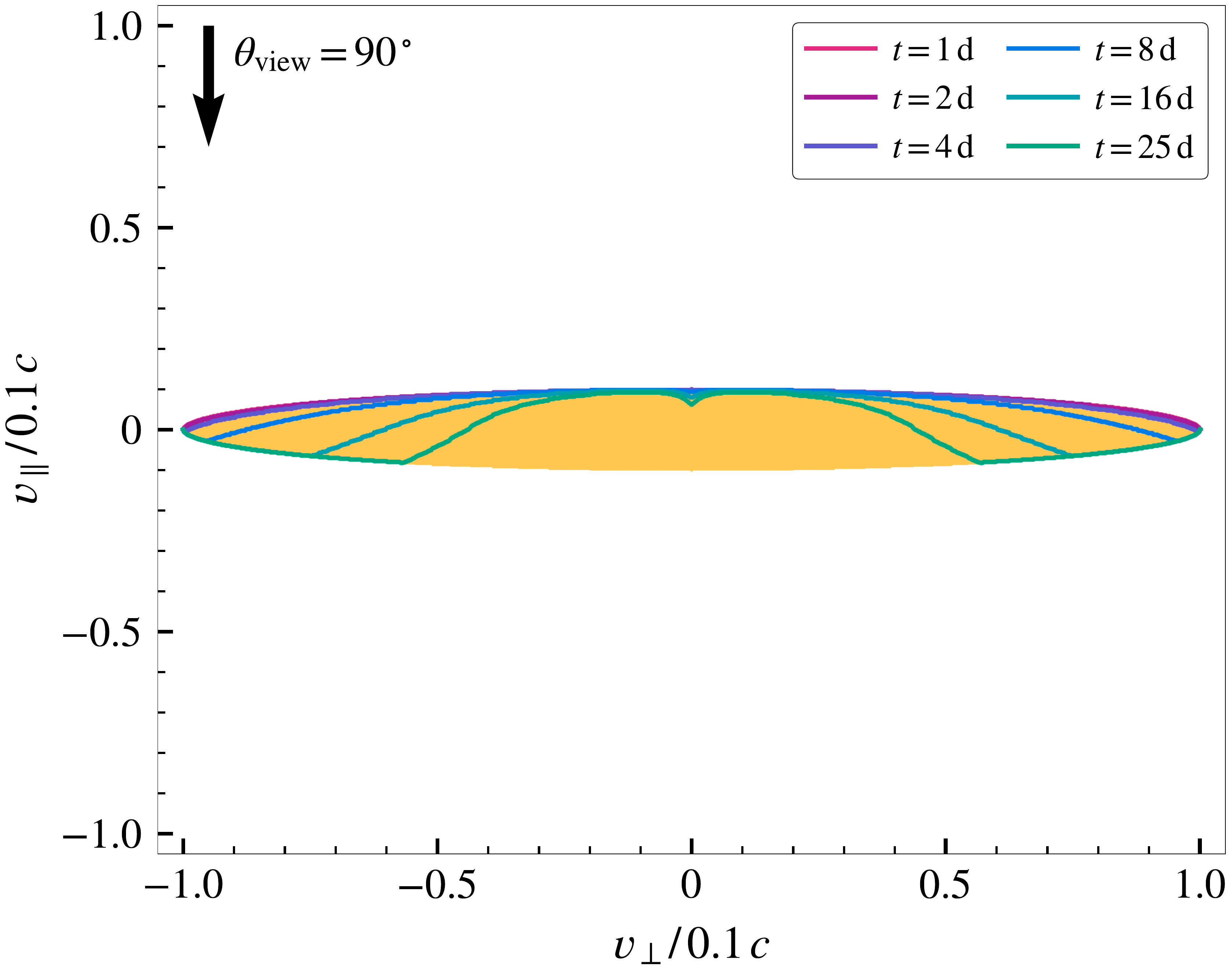}
    \caption{Sectional views of the photosphere evolution in velocity space of our fiducial model, as seen by observers at different viewing angles. The line of sight is fixed to be vertically downward in the planes of the figure, and only the cross-section through the ejecta center is shown. Here, $\varv_{\parallel}$ and $\varv_{\perp}$ represent the velocity components parallel and perpendicular to the line of sight, respectively. Colored lines indicate the photospheric contours at various epochs following the WD--NS merger.} 
    \label{ fig:photosphere }
\end{figure*}

Fig.~\ref{ fig:photosphere } shows the 2D sectional views of the photospheric evolution in velocity space for the fiducial model, as seen from different viewing angles. The central slices (i.e., cross-sections through the center of the ejecta) are taken, with the line of sight fixed vertically downward. At early times (${t \lesssim 4\,\mathrm{d}}$), the observed photosphere is located near the outermost layer of the ejecta, regardless of the viewing angle. Owing to the ellipsoidal velocity profile~(see Eq.~(\ref{ eq:v^max_w })), it is evident that the projected area of the observed photosphere naturally appears smaller for observers with lower $\theta_{\text{view}}$ values. As the ejecta expands, the photosphere gradually recedes into its inner layers. The morphology of the observed photosphere varies with the viewing angle. Compared to an on-axis observer ($\theta_{\text{view}} \simeq 0^\circ$), Fig.~\ref{ fig:photosphere } shows that off-axis observers can probe deeper, more central regions of the ejecta.


\section{Viewing-angle-dependent spectra and lightcurves}
\label{ sec:lc_spec }

Regarding the lightcurve construction, it is convenient to perform interpolation and integration by directly projecting the entire photosphere onto the mesh grid plane (see Sect.~\ref{ sec:photosphere })\footnote{This approach is well justified when the angular extent of the WD--NS merger is effectively unresolved by an observer at a cosmological distance.}. Accordingly, the flux at an observed frequency $\nu_{\rm obs}$ can be expressed as
\begin{equation}
\begin{aligned}
F_{\nu}\left(\nu_{\mathrm{obs}}\right) 
& \simeq \frac{2  }{h^{2} c^{2} D_{\mathrm{L}}^{2}} \iint_{S} \frac{\mathcal{D}(h \nu_{\rm obs} / \mathcal{D})^{3}}{\exp \left(h \nu_{\rm obs} / \mathcal{D} k_{\mathrm{B}} T_{\text {phot}}^{i j}\right)-1} d \sigma^{i j}\\
& = \frac{2  h}{c^{2} D_{\mathrm{L}}^{2}} \iint_{S} \frac{\nu_{\rm obs} ^ 3 / \mathcal{D}^{2}}{\exp \left(h \nu_{\rm obs} / \mathcal{D} k_{\mathrm{B}} T_{\text {phot}}^{i j}\right)-1} d \sigma^{i j} \,,
\end{aligned}
\label{ eq:flux_nu }
\end{equation}
where $h$ is the Planck constant, $D_{\mathrm{L}}$ is the luminosity distance to the source, and $k_{\mathrm{B}}$ is the Boltzmann constant, respectively. Defining the normalized wind velocity ${\beta \equiv \varv/ \it c}$, the Doppler factor is given by ${\mathcal{D}=\sqrt{1-\beta^{2}} /(1+\beta \cos \Delta \theta)}$, where $\Delta \theta$ is the angle between the velocity vector of the photosphere and the line of sight. The quantity $d \sigma^{i j}$ represents the infinitesimal projected photospheric surface element on the mesh grid. Note that flux densities per unit frequency and per unit wavelength are related by $F_{\nu} |d \nu|=F_{\lambda} |d \lambda|$, which leads to
\begin{equation}
\begin{aligned}
F_{\lambda}\left(\lambda_{\mathrm{obs}}\right) 
& \simeq \frac{2}{h^{4} c^{3} D_{\mathrm{L}}^{2}} \iint_{S} \frac{\mathcal{D}^{3}(h c / \mathcal{D} \lambda_{\mathrm{obs}})^{5}}{\exp \left(h c / \mathcal{D} k_{\mathrm{B}} \lambda_{\mathrm{obs}} T_{\text {phot}}^{i j}\right)-1} d \sigma^{i j}\\
& = \frac{2 h c^{2}}{ D_{\mathrm{L}}^{2}} \iint_{S} \frac{ 1 / \mathcal{D}^{2} \lambda_{\mathrm{obs}}^{5}}{\left[\exp \left(h c / \mathcal{D} k_{\mathrm{B}} \lambda_{\mathrm{obs}} T_{\text {phot}}^{i j}\right)-1 \right]} d \sigma^{i j} \,.
\end{aligned}
\label{ eq:flux_lambda }
\end{equation}
Furthermore, based on the calculated flux density $F_{\nu}$, one can derive the monochromatic AB magnitude using the standard definition $M_{\nu}=-2.5 \log _{10}\left(F_{\nu} / 3631\,\mathrm{Jy}\right)$. In all subsequent calculations, we adopt a luminosity distance of $D_{\mathrm{L}} = 10\,\mathrm{pc}$, so that the resulting magnitudes correspond to absolute magnitudes. Other distances can be simply rescaled.

\begin{figure*}[htbp]
  \centering
  \includegraphics[width=8cm]{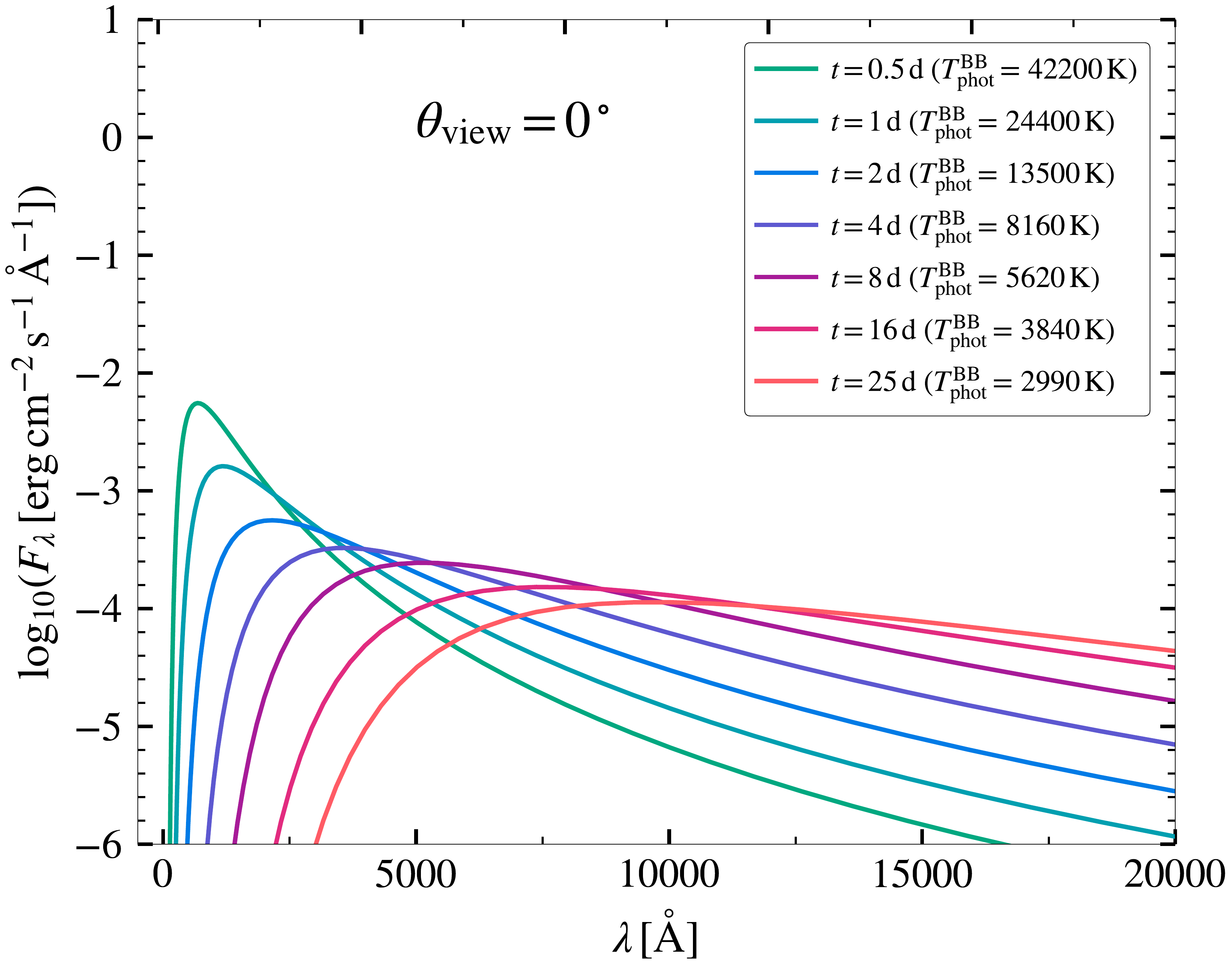}
  \hspace{1cm}
  \includegraphics[width=8cm]{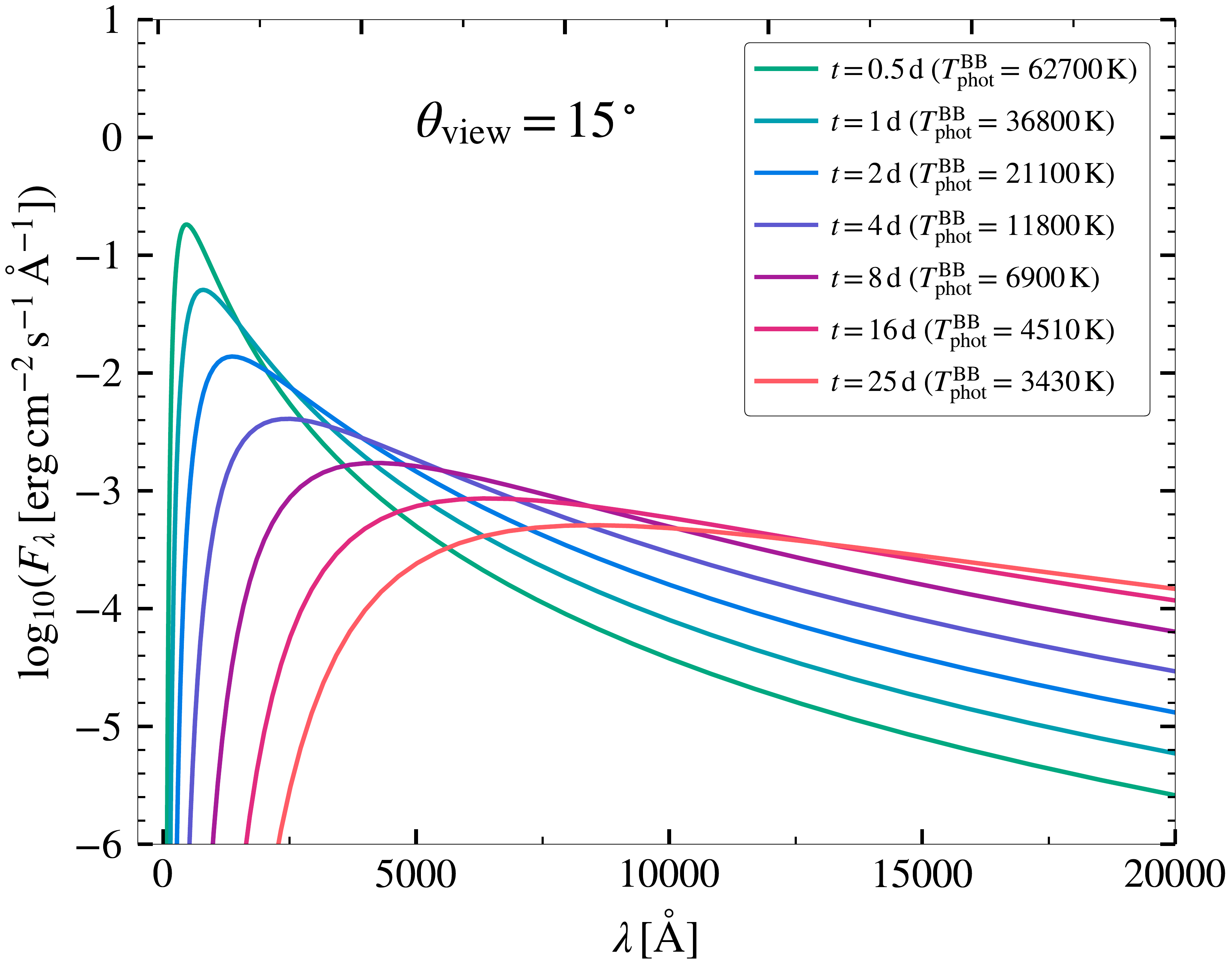}
  
  \vspace{0.3cm}
  
  \includegraphics[width=8cm]{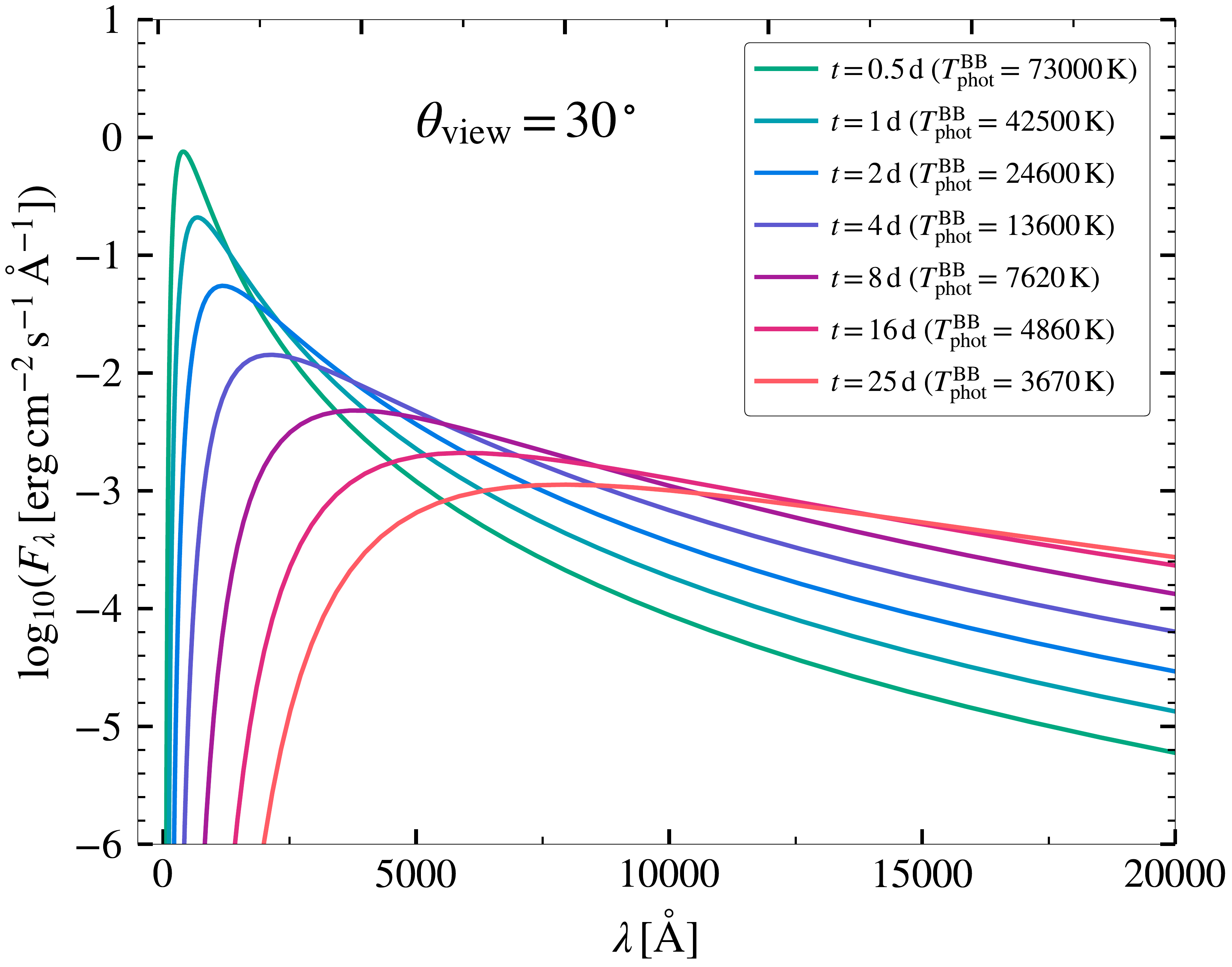}
  \hspace{1cm}
  \includegraphics[width=8cm]{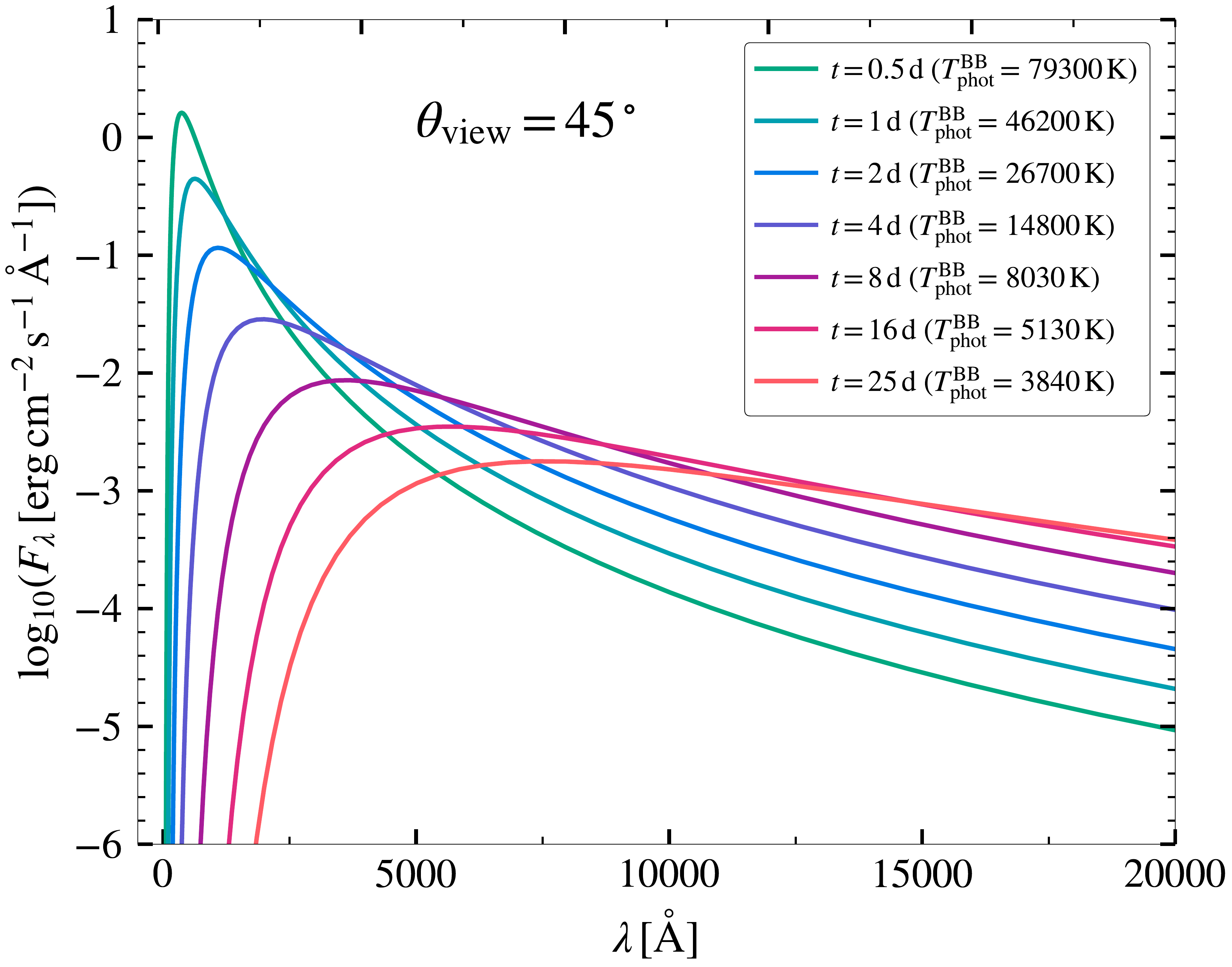}
  
  \vspace{0.3cm}
  
  \includegraphics[width=8cm]{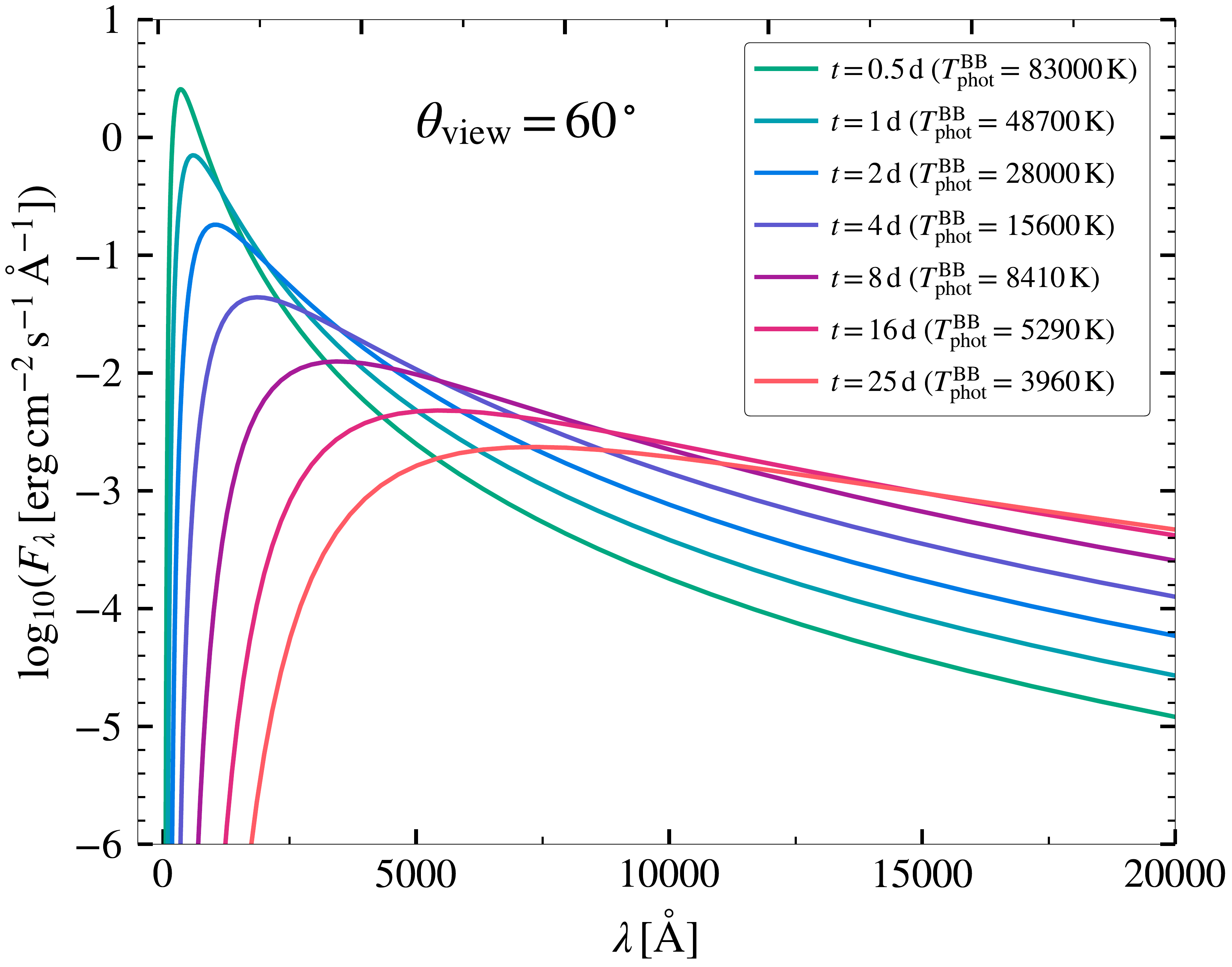}
  \hspace{1cm}
  \includegraphics[width=8cm]{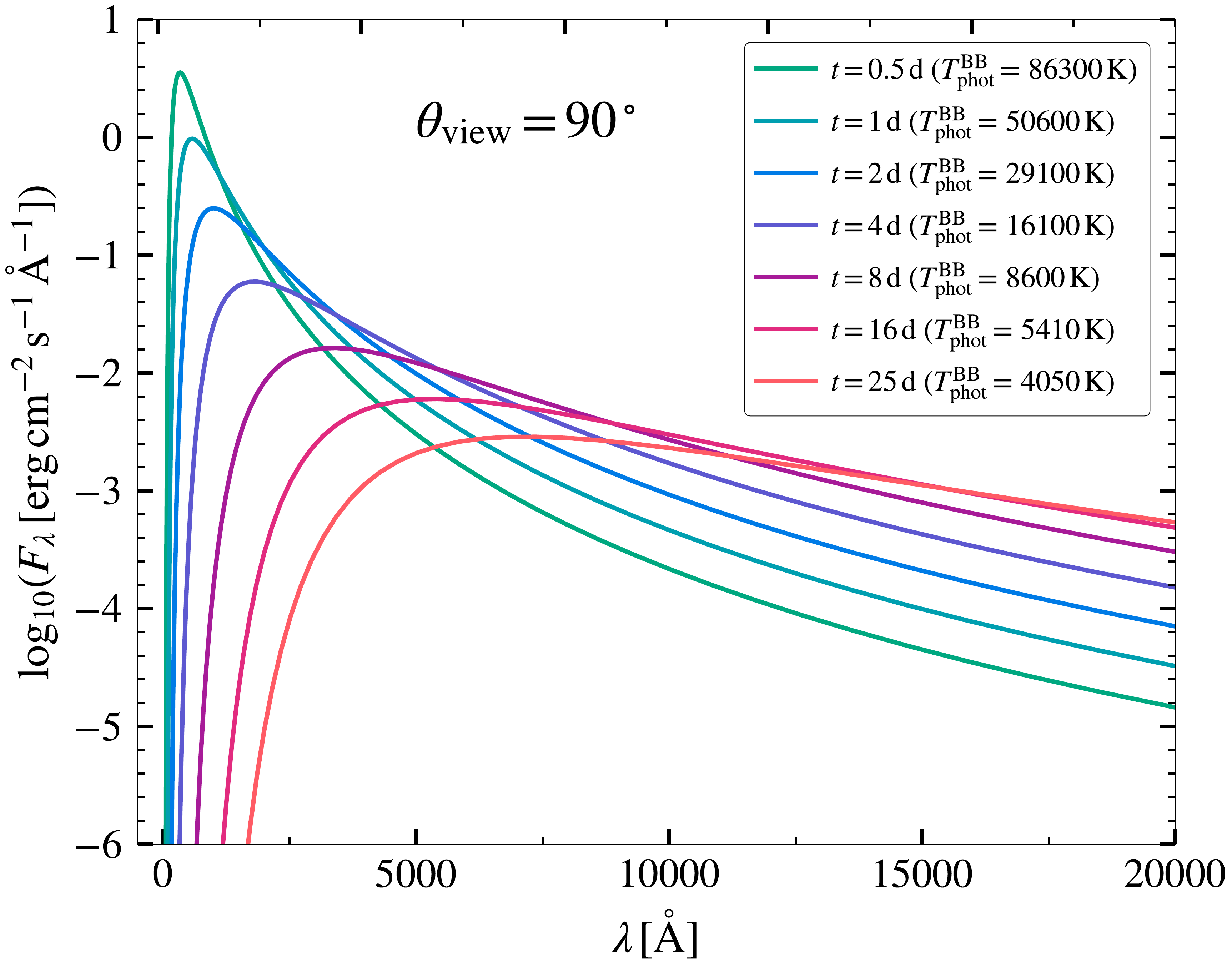}
    \caption{Emergent spectra of WD--NS mSNe for $\theta_{\text{view}} = 0^{\circ}$, $15^{\circ}$, $30^{\circ}$, $45^{\circ}$, $60^{\circ}$, and $90^{\circ}$. Colored lines show the total observed emergent spectra at different epochs after the WD--NS merger of our fiducial model, assuming a luminosity distance of $D_{\mathrm{L}} = 10\,\mathrm{pc}$. The derived photospheric blackbody temperatures, $T^{\rm BB}_{\rm phot}$, are shown in the labels.} 
    \label{ fig:flux }
\end{figure*}

\begin{figure*}[htbp]
  \centering
  \includegraphics[width=8cm]{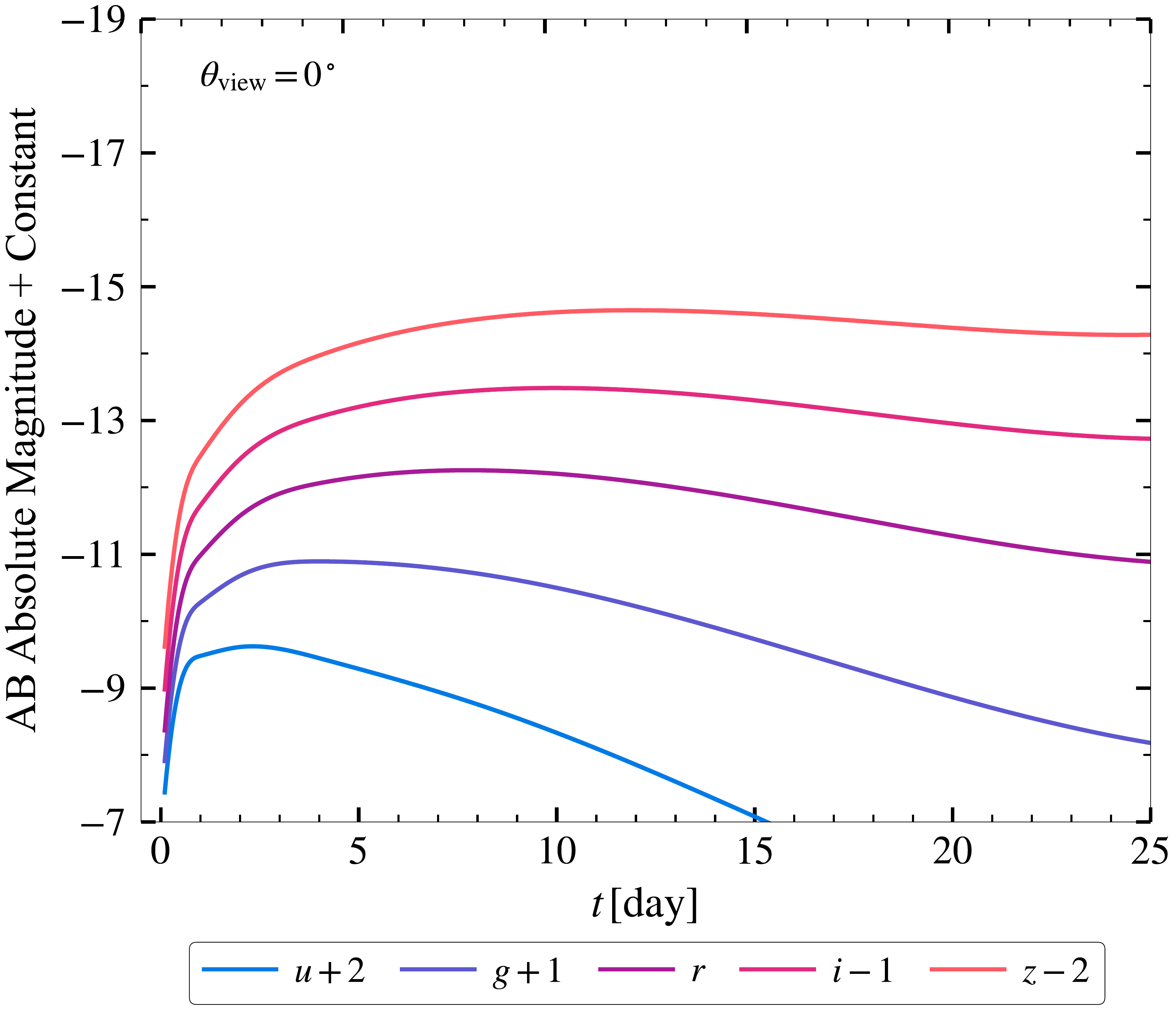}
  \hspace{1cm}
  \includegraphics[width=8cm]{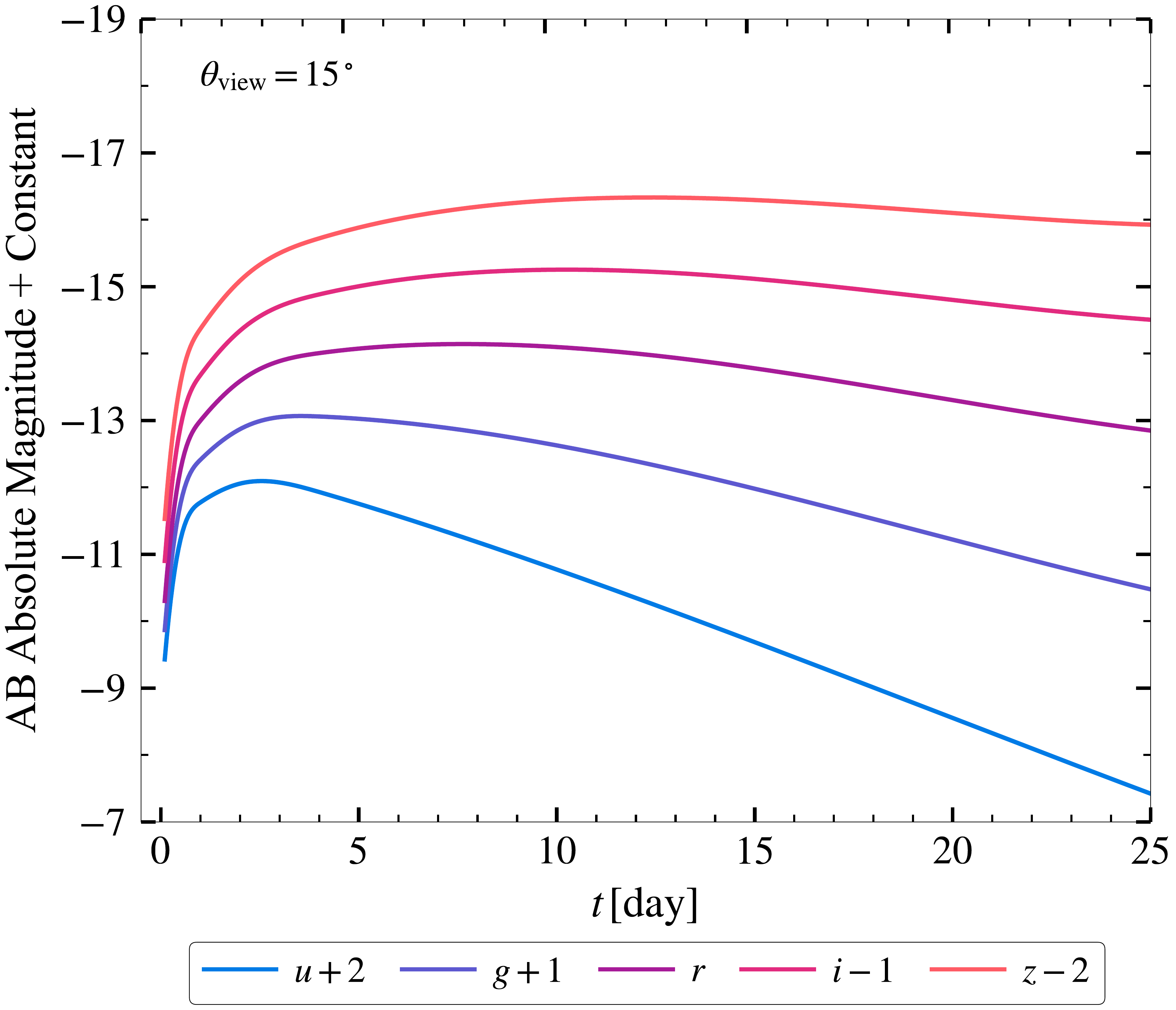}
  
  \vspace{0.3cm}
  
  \includegraphics[width=8cm]{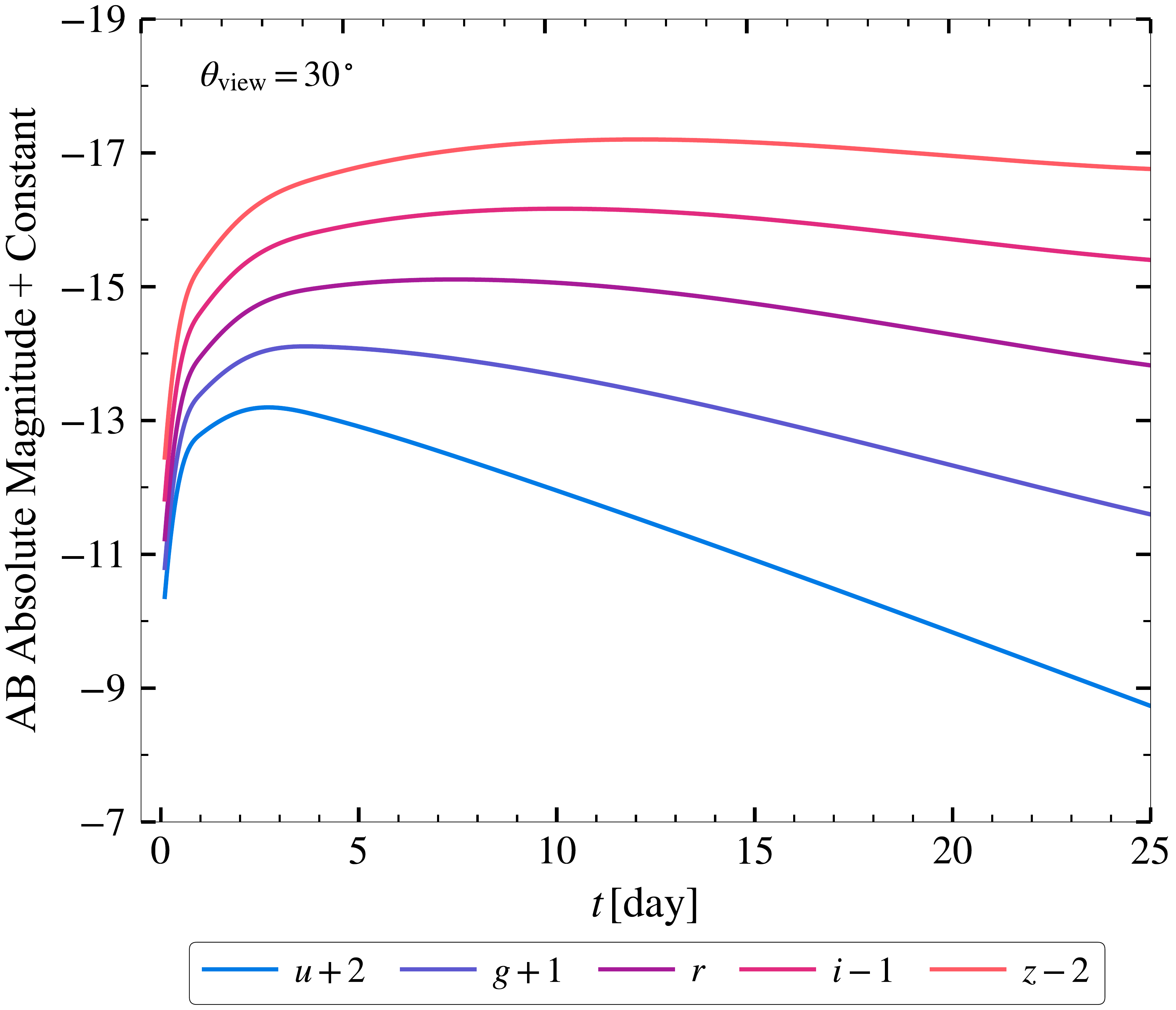}
  \hspace{1cm}
  \includegraphics[width=8cm]{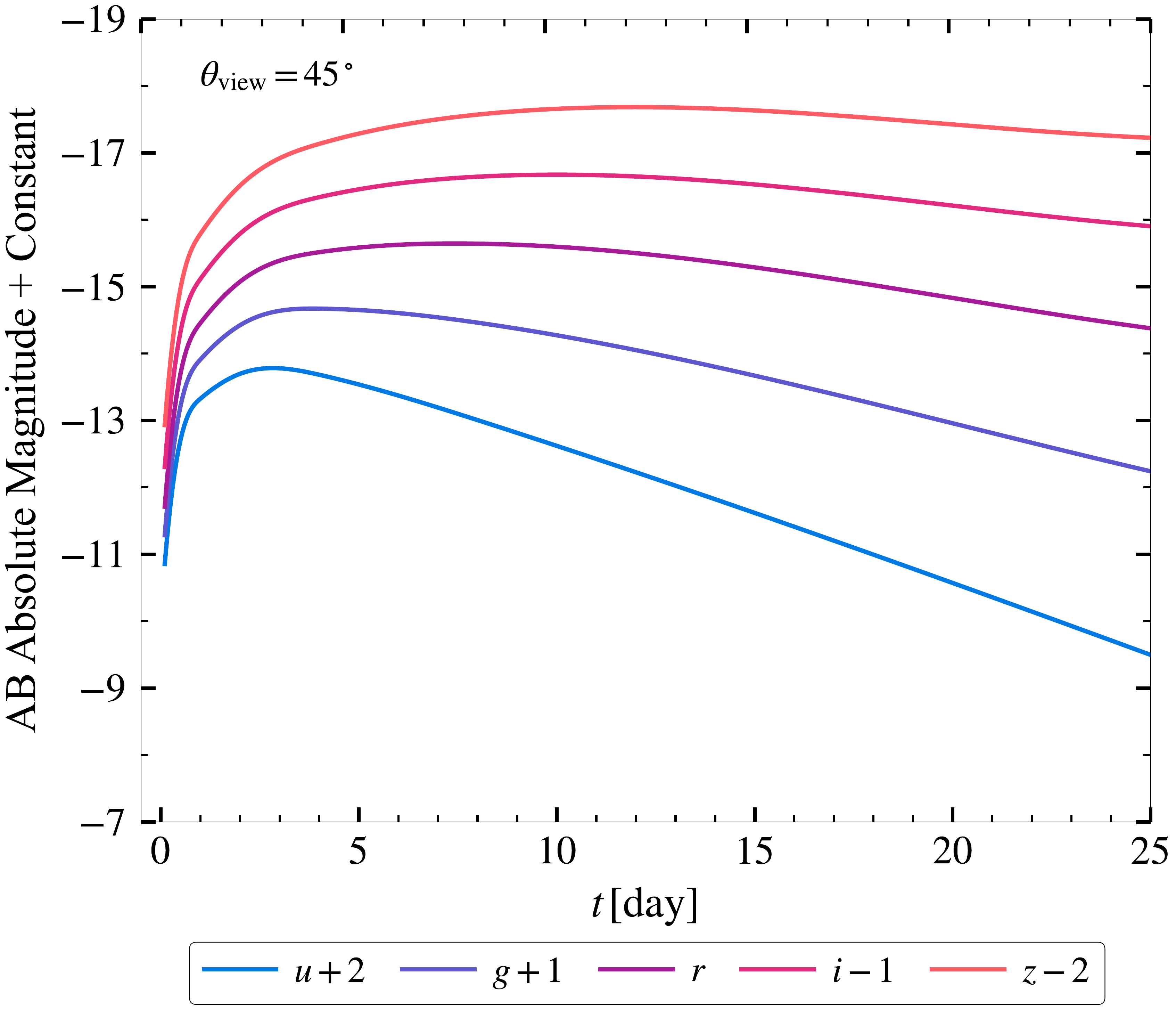}
  
  \vspace{0.3cm}
  
  \includegraphics[width=8cm]{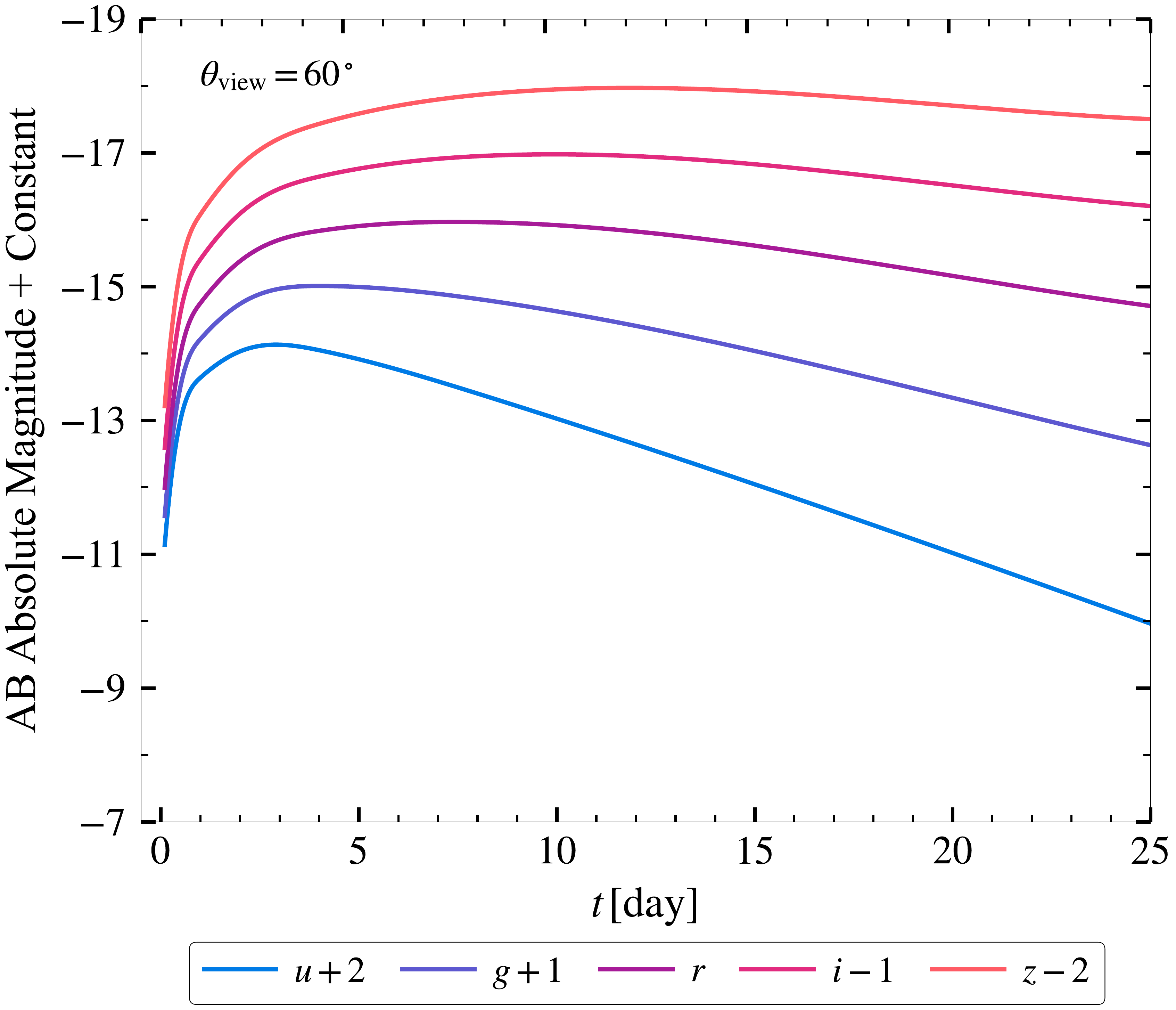}
  \hspace{1cm}
  \includegraphics[width=8cm]{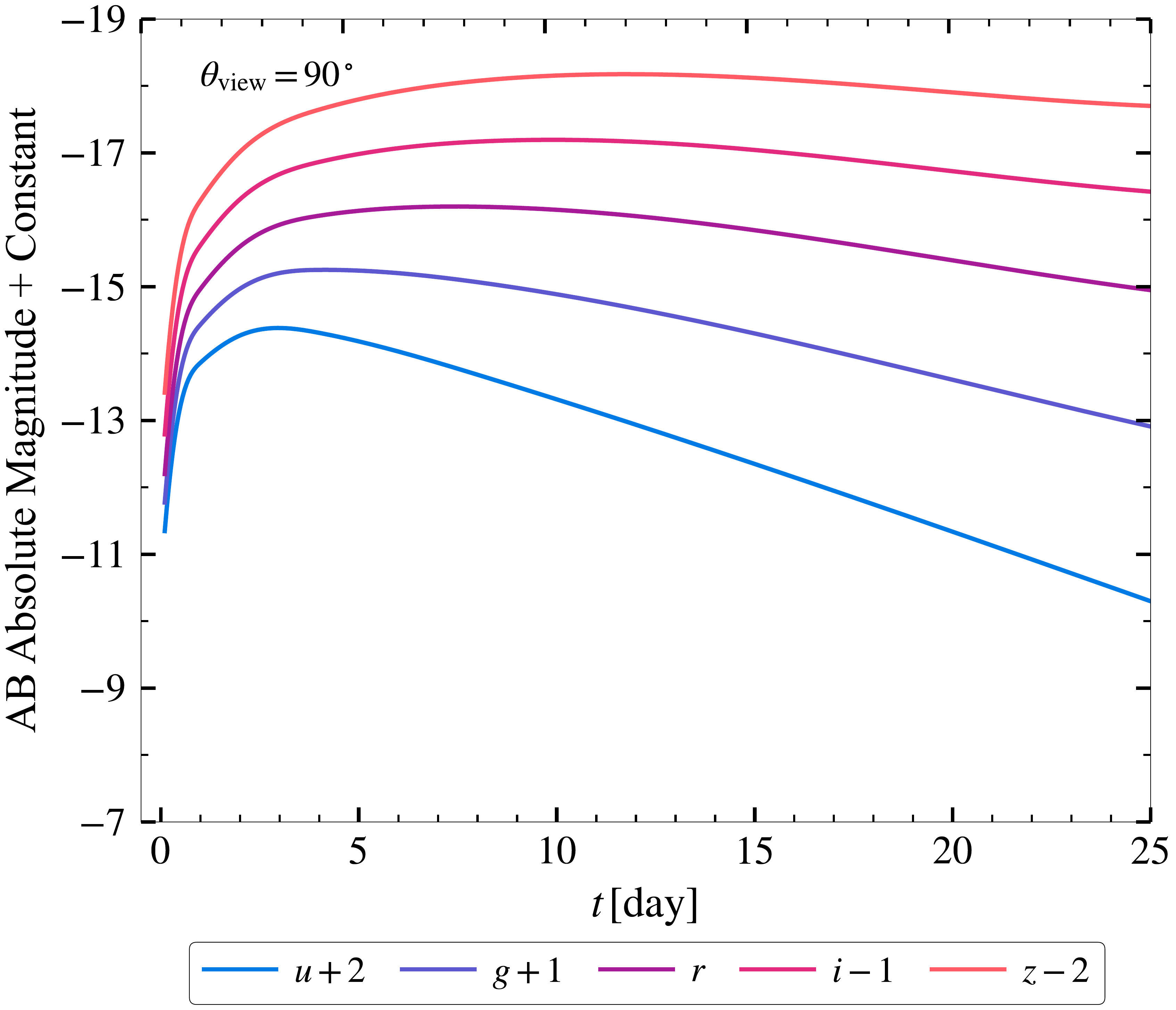}
    \caption{Viewing-angle-dependent $ugriz$ lightcurves of WD--NS mSNe for the fiducial model. Six different viewing angles $\theta_{\rm view}$ are considered: $0^\circ$ (top left), $15^{\circ}$ (top right), $30^{\circ}$ (middle left), $45^{\circ}$ (middle right), $60^{\circ}$ (bottom left), and $90^{\circ}$ (bottom right).} 
    \label{ fig:magnitude }
\end{figure*}

\begin{figure*}[htbp]
\sidecaption
\includegraphics[width=12cm]{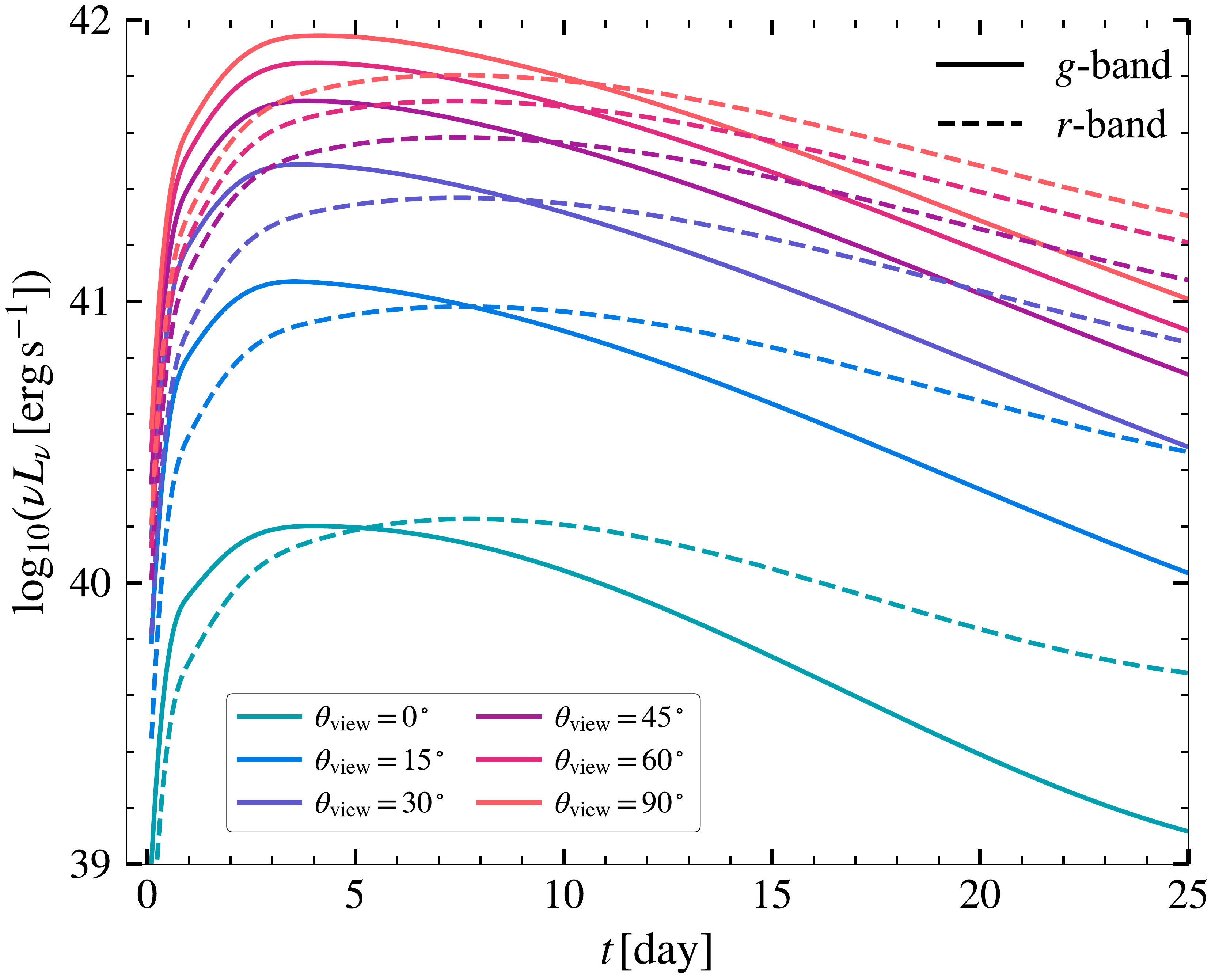}
\caption{Optical luminosity lightcurve of WD--NS mSNe for our fiducial model. Different colors indicate different viewing angles at $\theta_{\text{view}} = 0^{\circ}$, $15^{\circ}$, $30^{\circ}$, $45^{\circ}$, $60^{\circ}$, and $90^{\circ}$. Solid and dashed lines correspond to the $g$- and $r$-band luminosities, respectively.}
\label{ fig:luminosity }
\end{figure*}

\begin{figure*}[htbp]
\sidecaption
\includegraphics[width=12cm]{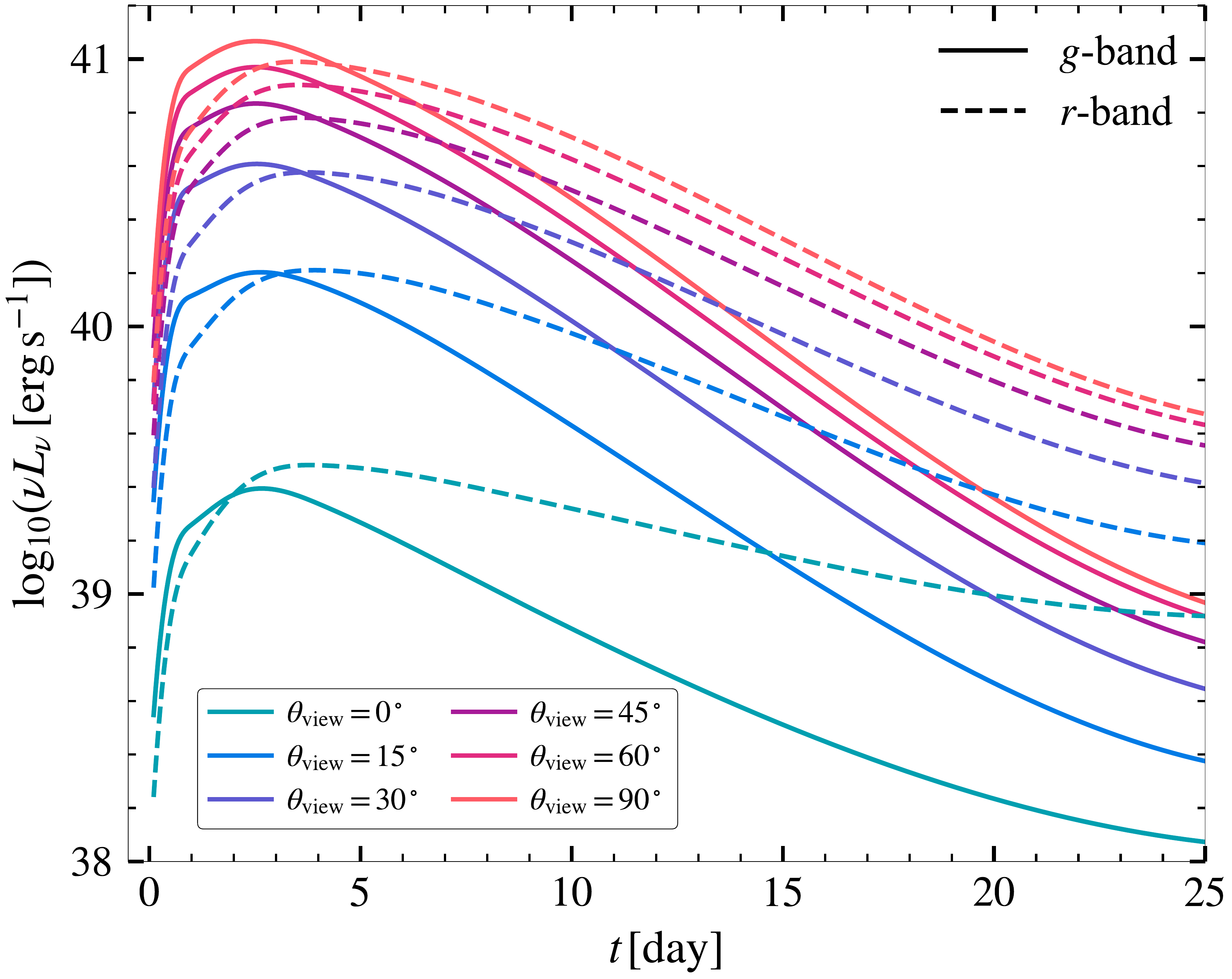}
\caption{Same as Fig.~\ref{ fig:luminosity }, but assuming an ejecta mass of ${M_\mathrm{ej} = 0.03\,\mathrm{M}_{\odot}}$ and a $^{56}$Ni mass of ${M_\mathrm{Ni} = 0.001\,\mathrm{M}_{\odot}}$.}
\label{ fig:luminosity_lower }
\end{figure*}

\begin{figure*}[htbp]
\centering
\includegraphics[width=17cm]{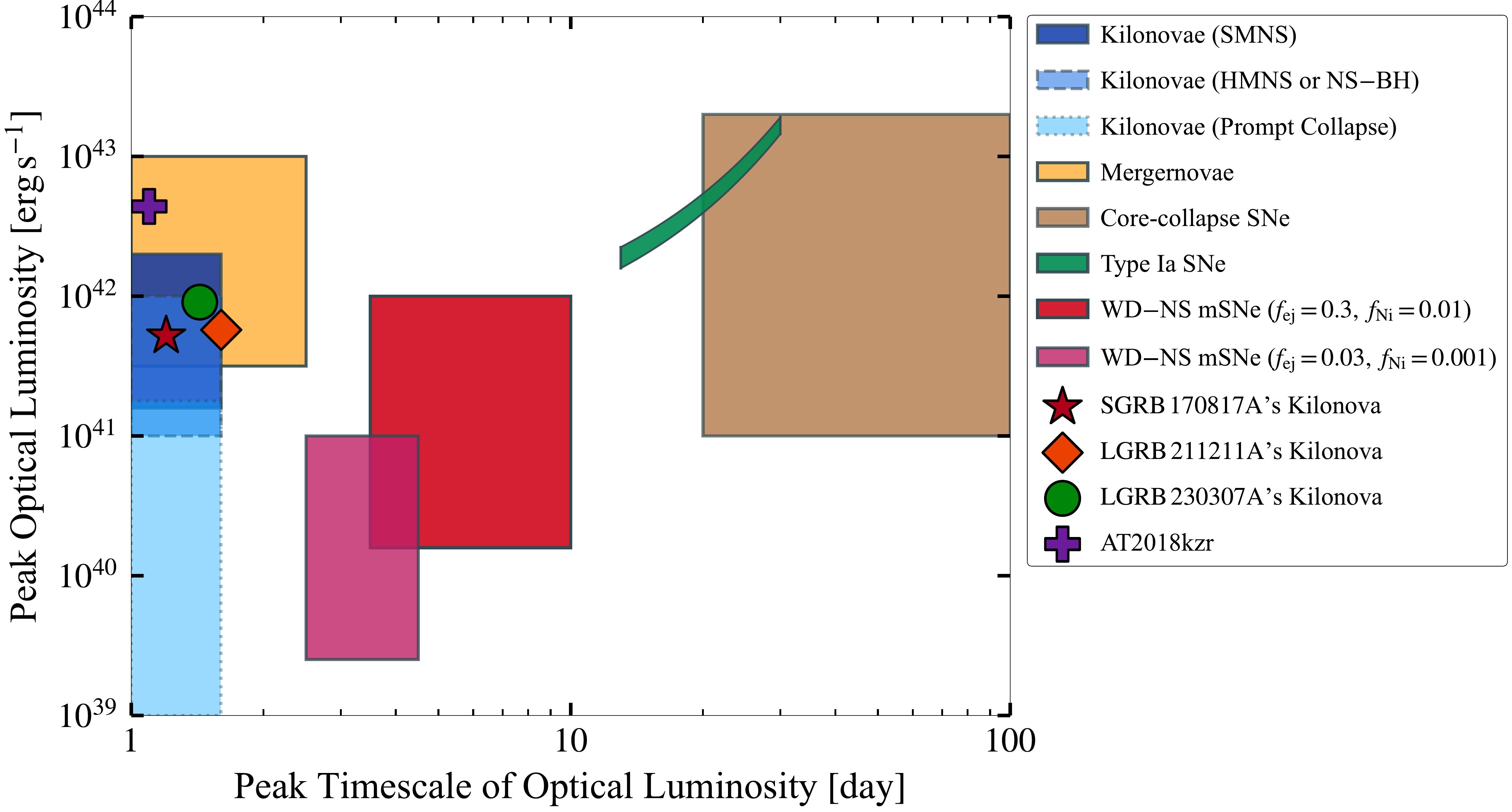}
\caption{Comparisons of the peak timescale and peak optical luminosity among various optical transients for ${t \gtrsim 1\, \mathrm{ d }}$. The three blue-shaded regions, from dark to light, correspond to kilonova models associated with SMNS, HMNS or NS--BH, and prompt collapse scenarios, respectively \citep{2020ApJ...889..171K}. The yellow-shaded region indicates the parameter space for central-engine-powered kilonovae \citep[i.e., the so-called ``mergernova'' model;][]{Yu:2013kra, Ai:2024zcn}. The brown and green shaded regions denote the parameter spaces of core-collapse and type Ia SNe, respectively \citep{Kasliwal:2012gfm, 2017NatAs...1E...8C, Inserra:2019ciq}. The red and pink shaded regions represent the WD--NS mSNe explored in this work, corresponding to models with fiducial and reduced values of $f_{\mathrm{ej}}$ and $f_{\mathrm{Ni}}$, respectively. For reference, we also indicate three well-known kilonova events associated with SGRB\,170817A \citep[red star;][]{Villar:2017wcc}, LGRB\,211211A \citep[orange diamond;][]{Yang:2022qmy}, and LGRB\,230307A \citep[green circle][]{JWST:2023jqa}, as well as the candidate WD--NS merger event AT2018kzr \citep[purple cross;][]{McBrien:2019wfv}.}
\label{ fig:comparison }
\end{figure*}

Figure~\ref{ fig:flux } shows the temporal evolution of the emergent spectra from the WD--NS ejecta of the fiducial model at various viewing angles. Regardless of the viewing angle, the peak emission initially lies in the ultraviolet regime for $t \lesssim 4\,\mathrm{d}$, and gradually shifts to the optical and then to the infrared bands with time. The photospheric blackbody temperature can be estimated from Wien's law, i.e., $T^{\rm BB} = b/\lambda_{\rm max}$, where $b \simeq 0.29\,{\rm cm}\,{\rm K}$ is the Wien's displacement constant and $\lambda_{\max }$ is the wavelength corresponding to the peak radiation intensity. As shown in Fig.~\ref{ fig:flux }, both the spectral intensity and the photospheric blackbody temperature exhibit significant viewing-angle dependence. For on-axis observers, $T^{\rm BB}_{\rm phot}$ cools from $\simeq 4.2 \times 10^{4}\,\mathrm{K}$ at $t = 0.5\,\mathrm{d}$ to $\simeq 3 \times 10^{3}\,\mathrm{K}$ at $t = 25\,\mathrm{d}$; whereas for equatorial observers, it decreases from $\simeq 8.6 \times 10^{4}\,\mathrm{K}$ to $\simeq 4 \times 10^{3}\,\mathrm{K}$ over the same time interval. Observers located at larger viewing angles receive substantially higher flux densities---up to nearly two orders of magnitude larger at early times. This enhancement arises from the emergence of more centrally located (and hotter) photospheric layers at larger viewing angles, as well as the increased projected photospheric area, as discussed in Sect.~\ref{ sec:photosphere }.

To further illustrate the viewing-angle dependence of the multiband emission following a WD--NS merger of our fidiucial model, we present $ugriz$-band\footnote{Details of the Sloan filter system can be found at \url{https://www.sdss4.org}.} lightcurves in Fig.~\ref{ fig:magnitude } for different viewing angles. We find that the peak absolute magnitudes in the optical bands range from ${\simeq -12\,\mathrm{mag}}$ for an on-axis view ($\theta_{\text{view}} \simeq 0^\circ$) to ${\simeq -16\,\mathrm{mag}}$ for an off-axis view ($\theta_{\text{view}} \simeq 90^\circ$). These optical peak luminosities are comparable to those of kilonovae from BNS and NS--BH mergers \citep[see e.g.,][]{Nakar:2019fza, 2020ApJ...889..171K, Metzger:2019zeh}, though WD--NS thermal transients typically exhibit longer timescales to peak, usually $\gtrsim 3$--$10\,\mathrm{d}$. Since these WD--NS merger transients are powered solely by $^{56}\mathrm{Ni}$ heating, we introduce the term ``mini-supernovae'' (mSNe) to describe their thermal emission. Using the time-dependent emergent spectra obtained for WD--NS mSNe, we also calculate the luminosity lightcurves via $\nu L_{\nu}(t) = 4 \pi D_{\mathrm{L}}^{2} \nu F_{\nu}$, where $F_{\nu}$ is given by Eq.~(\ref{ eq:flux_nu }). Fig.~\ref{ fig:luminosity } shows the optical ($g$ and $r$ band) luminosity evolution for different viewing angles. The WD--NS merger can power mSNe with peak luminosities in the range $\simeq 10^{40}$--$10^{42}\,\mathrm{erg} \, \mathrm{s}^{-1}$, depending on the viewing geometry. The peak timescale typically falls within \mbox{$3$--$10\,\mathrm{d}$}. Note that the observed directions $\theta_{\text{view}} \simeq 0^\circ$ and $\theta_{\text{view}} \simeq 90^\circ$ correspond to extreme and relatively rare cases, most events are expected to be viewed from intermediate angles, yielding typical peak optical luminosities of ${\simeq 10^{41}\,\mathrm{erg} \, \mathrm{s}^{-1}}$ with a peak timescale of $3$--$10\,\mathrm{d}$. 

We also investigate an alternative parameter set with ${f_{\mathrm{ej}} = 0.03}$ and $f_{\mathrm{Ni}} = 0.001$, corresponding to an ejecta mass of $M_{\rm ej}=0.03\,\mathrm{M}_\odot$ and a $^{56}$Ni mass of $M_{\rm Ni}=0.001\,\mathrm{M}_\odot$, both one order of magnitude lower than those adopted in the fiducial model. This case is more relevant to stellar-mass BH-WD or He WD--NS mergers, which share similar ejecta properties but typically produce smaller ejecta and $^{56}$Ni masses \citep[e.g.,][]{Metzger:2011zk, Fernandez:2012kh, Margalit:2016joe, Zenati:2018gcp, Fernandez:2019eqo, Bobrick:2021hho}. The resulting optical lightcurves are shown in Fig.~\ref{ fig:luminosity_lower } along different lines of sight. We find that the WD--NS mSNe with ejecta and $^{56}$Ni masses reduced by one order of magnitude compared to the fiducial model are roughly an order of magnitude fainter, with peak luminosities of $\simeq 10^{39}$--$10^{41}\,\mathrm{erg} \, \mathrm{s}^{-1}$ and shorter peak timescales of $2$--$5\,\mathrm{d}$.


\section{Discussion}
\label{ sec:discussion }

In comparison, our results show good consistency with previous studies on WD--NS transients. For example, based on 1D steady-state accretion disk models for WD--NS mergers, \citet{Metzger:2011zk} suggested that a large fraction ($\gtrsim 50\%$) of the WD mass can be unbound as ejecta, simultaneously synthesizing $^{56}$Ni masses of $10^{-3}$--$10^{-2} \, \mathrm{M}_{\odot}$ inside the ejecta. By treating the unbound ejecta as a single homologous shell and neglecting angular distributions, \citet{Metzger:2011zk} predicted optical transients with week-long peak timescales and peak luminosities of ${\simeq 10^{39}}$--${10^{41.5}\,\mathrm{erg} \, \mathrm{s}^{-1}}$ following the semianalytical model of \citet{Kulkarni:2005jw}. Using $^{56}$Ni and ejecta velocity distributions from numerical simulations of CO and ONe WD accretion disks, and neglecting angular corrections, \citet{Fernandez:2019eqo} showed that WD--NS merger transients could achieve peak luminosities of $\simeq 10^{40}$--$10^{41}\,\mathrm{erg} \, \mathrm{s}^{-1}$ with week-long peak timescales inferred by the Arnett's semi-analytical model \citep{Arnett1979} as well. \citet{Zenati:2019glf} performed 3D hydrodynamic (\texttt{SPH}) and 2D hydrodynamic-thermonuclear (\texttt{FLASH}) to study the disruption of CO WDs by NSs. They subsequently post-processed the simulations with a large nuclear reaction network and applied the \texttt{SuperNu} radiation-transfer code, assuming homologous ejecta expansion, to predict the observational signatures and detailed properties of CO WD--NS merger transients. Their models yielded faint, red transients with peak luminosities around ${10^{41}\,\mathrm{erg} \, \mathrm{s}^{-1}}$ and peak timescales ranging from weeks to months, while neglecting viewing-angle dependence. Using a 1D analytic advection-dominated accretion-disk model, \citet{Kaltenborn:2022vxf} systematically investigated the inflow and outflow structure of WD--NS post-merger disks, including the effects of nuclear burning, neutrino emission, and wind ejection. They further simulated multiband lightcurves powered by radioactive iron-group elements with the \texttt{SuperNu} radiation-transfer code, predicting optical transients peaking at \mbox{$\simeq 10^{41}$--$10^{42}\,\mathrm{erg} \, \mathrm{s}^{-1}$} with $5$--$8\,\mathrm{d}$ peak timescales. More recently, \citet{Liu:2025voz} employed detailed nucleosynthesis calculations (\texttt{SkyNet}) in spherically symmetric ejecta models for WD--NS mergers, predicting optical transients with peak luminosities of a few ${10^{41}\,\mathrm{erg} \, \mathrm{s}^{-1}}$ and peak timescales of $2$--$4\,\mathrm{d}$. Overall, while our results are broadly consistent with previous studies in terms of peak luminosities and timescales, our work uniquely highlights the previously unexplored viewing-angle dependence of WD--NS merger transients.

We then explore how our WD--NS mSNe fit within the broader landscape of various optical transients in terms of their peak optical luminosities and timescales. Most importantly, Fig.~\ref{ fig:comparison } compares different kilonova scenarios, including those from BNS and NS--BH mergers, to illustrate where WD--NS mSNe lie in terms of their peak luminosities and timescales. Among these, \citet{2020ApJ...889..171K} investigated the post-merger evolution and typical ejecta properties for both BNS and NS--BH binaries\footnote{Due to temperature calculation limitations, the peak brightness in \citet{2020ApJ...889..171K} corresponds to the brightest point at ${t \gtrsim 1\, \mathrm{ d }}$. We adopt the same convention in Fig.~\ref{ fig:comparison }, but notice that BNS and NS--BH mergers could yield even brighter emission at earlier times, particularly in the ultraviolet band.}. For BNS systems in particular, they considered different types of merger remnants, including a long-lived supermassive NS (SMNS), a hypermassive NS (HMNS), and a promptly-collapsed BH. These models predict that BNS and NS--BH kilonovae peak significantly earlier than the WD--NS counterparts presented in this work. Additionally, if a millisecond magnetar survives the BNS merger and can injects extra energy into the ejecta \citep[see e.g.,][]{Yu:2013kra,Metzger2014, Ai:2024zcn}, the resulting kilonova lightcurve can become both brighter and longer‐lived than that of a standard BNS kilonova. In contrast to typical kilonovae from BNS or NS--BH mergers, our WD--NS mSNe occupy the lower-right region of the parameter space, meaning that they are comparatively dimmer and peak significantly later. When adopting reduced values of $f_{\mathrm{ej}}$ and $f_{\mathrm{Ni}}$, the region occupied by WD--NS mSNe shifts slightly toward the lower-left, indicating somewhat fainter and faster-evolving events compared to those with our fiducial $f_{\mathrm{ej}}$ and $f_{\mathrm{Ni}}$ values. 

Fig.~\ref{ fig:comparison } also highlights three well-known kilonova events associated with SGRB\,170817A \citep{Villar:2017wcc}, LGRB\,211211A \citep{Yang:2022qmy}, and LGRB\,230307A \citep{JWST:2023jqa}, all of which are consistent with BNS or NS--BH merger scenarios and are difficult to explain with our $^{56}$Ni-powered WD--NS mSNe. Notably, if the WD--NS system contains a massive WD near the Chandrasekhar mass limit, the NS may merge into the centre of the WD and trigger its collapse, potentially forming a millisecond magnetar as the central engine \citep{Yang:2022qmy, Cheong:2024hrd, Zhang:2024syc}. Alternatively, however, some recent studies suggested that in WD--NS mergers, the NS may be spun up by accreting disrupted WD material and its magnetic field can be amplified via dynamo processes, finally leaving behind a millisecond magnetar remnant \citep[see e.g.,][]{2020ApJ...893....9Z, Kaltenborn:2022vxf, Zhong:2023zwh, Zhong:2024alw,Wang:2024ijo}. Collectively, these studies offer valuable insights into the possible formation of a magnetar central engine and the corresponding central-engine-powered optical transients following WD--NS mergers, in which the ejecta can be accelerated and heated by magnetar spin-down, leading to emission with shorter peak timescales and higher peak luminosities than those of our $^{56}$Ni-powered WD--NS mSNe. As shown in Fig.~\ref{ fig:comparison }, kilonovae associated with merger-origin LGRBs (i.e., LGRB\,211211A and LGRB\,230307A), may still possibly arise from WD--NS mergers, but this requires the formation of a millisecond magnetar as the central engine to power them. Similar modeling assuming a WD--NS merger origin with a magnetar remnant can also account for other fast-evolving optical transients, such as AT2018kzr \citep[see Fig.~\ref{ fig:comparison };][]{McBrien:2019wfv, Gillanders:2020fhm}. We note that the ultimate fate of the merger remnant in WD--NS systems remains uncertain, and the probability of forming such magnetar-powered events in WD--NS mergers has yet to be conclusively determined. Since WD--NS mergers are generally not expected to produce significant amounts of $r$-process material, the spectroscopic identification of intermediate-mass elements (e.g., O, Mg, Si, and Ca) and iron-peak elements in associated mSNe would strongly motivate a WD--NS merger origin \citep{McBrien:2019wfv, Gillanders:2020fhm, Kaltenborn:2022vxf, Liu:2025voz}. Future discoveries of additional WD--NS merger candidates will provide valuable insights into their evolutionary pathways and remnant outcomes. On the theoretical side, more realistic models incorporating detailed dynamical and thermodynamical evolution, as well as temperature- and wavelength-dependent opacities, are essential to fully understand these WD--NS transients, which we leave for future work.


\section{Conclusion}
\label{ sec:conclu }

Building on previous numerical simulations of the WD--NS binaries undergoing unstable mass transfer, we adopt a semi-analytical discretization approach to investigate the viewing-angle dependence of thermal emission from such mergers. Using a 2D axisymmetric polar-dominated wind ejecta model with heating from the radioactive decay of $^{56}$Ni, we track the evolution of the viewing-angle‐dependent photosphere and its temperature. We find that for a WD--NS merger with a typical ejecta mass of $M_{\rm ej}=0.3\,{\rm M}_\odot$ and a $^{56}$Ni mass of $M_{\rm Ni}=0.01\,{\rm M}_\odot$, the photosphere resides near the outermost ejecta layers for all viewing angles within ${\lesssim 4\,\mathrm{d}}$ after the merger, receding to deeper layers as the ejecta expands. The geometry of the observed photosphere varies with the viewing angle: off-axis observers see more centrally located, hotter ejecta regions than on-axis observers. Along the polar direction, the photospheric blackbody temperature cools from $\simeq 4.2 \times 10^{4}\,\mathrm{K}$ to $\simeq 3 \times 10^{3}\,\mathrm{K}$ between $0.5\,{\rm d}$ and $25\,{\rm d}$, while along the equatorial viewing it decreases from $\simeq 8.6 \times 10^{4}\,\mathrm{K}$ to $\simeq 4 \times 10^{3}\,\mathrm{K}$ over the same period. Our results show that the resulting emission intensity also has a strong viewing‐angle dependence: larger viewing angles yield significantly higher fluxes due to both a larger projected photospheric area and access to more centrally located (and hotter) ejecta layers. The peak absolute magnitudes in the optical bands range from ${\simeq -12\,\mathrm{mag}}$ at polar line of sight to ${\simeq -16\,\mathrm{mag}}$ at equatorial line of sight, corresponding to peak optical luminosities of \mbox{$\simeq 10^{40}$--$10^{42}\,\mathrm{erg} \, \mathrm{s}^{-1}$}. Thus, the peak optical luminosities of the thermal emission following WD--NS mergers are comparable to those of kilonovae from BNS and NS--BH mergers, though WD--NS thermal transients generally evolve more slowly, with typical peak timescales of $3$--$10\,\mathrm{d}$. Furthermore, lower ejecta masses and reduced $^{56}\mathrm{Ni}$ content naturally produce fainter transients with shorter peak timescales (see Fig.~\ref{ fig:comparison }). 

While our results are broadly consistent with earlier studies of WD--NS transients in terms of peak luminosities and peak timescales, our work uniquely emphasizes the viewing-angle dependence of WD--NS mSNe, which is not systematically addressed before. For kilonovae from BNS or NS--BH mergers, \citet{2020ApJ...889..171K} showed that the observed flux is typically higher in the polar direction due to the presence of multiple ejecta components, especially the optically thick dynamical ejecta concentrated in the equatorial plane. In contrast, WD--NS mergers are expected to exhibit a fundamentally different ejecta morphology. As highlighted by \citet{Bobrick:2021hho}, only a small fraction of WD debris (a few times $10^{-3}\,\mathrm{M}_\odot$) is expected to be dynamically ejected and unbound. This ejecta undergoes negligible nuclear processing to synthesize $^{56}\mathrm{Ni}$ \citep{Metzger:2011zk, Margalit:2016joe, Fernandez:2019eqo, Kaltenborn:2022vxf}, resulting in a negligible observational impact. Therefore, our work can inspire further investigations into the observational signatures and diversity of WD--NS merger transients. Moreover, given the rapid development in both ground-based and space-based GW astronomy, future multimessenger observations also have potential to provide more insights into WD--NS mergers \citep{Yin:2023gwc, Kang:2023rux, Moran-Fraile:2023oui, Zhong:2024sry, Du:2024kst}, helping to further constrain their properties and enrich our understanding of these systems.


\begin{acknowledgements}

We thank the anonymous referee for helpful comments. This work was supported by the National SKA Program of China (2020SKA0120300), the Beijing Natural Science Foundation (QY25099, 1242018), the National Natural Science Foundation of China (12573042), and the Max Planck Partner Group Program funded by the Max Planck Society. YK is supported by the China Scholarship Council (CSC).

\end{acknowledgements}


\bibliographystyle{aa}
\bibliography{refs} 


\end{document}